\documentclass[a4paper,11pt]{article}
\pdfoutput=1
\usepackage{geometry}
\usepackage{a4wide}
\usepackage{graphicx}
\usepackage{epsf}
\usepackage{amsmath}
\usepackage[normalem]{ulem}
\usepackage{amssymb}

\usepackage{cite}
\usepackage{multirow,tabularx}
\usepackage{appendix}
\usepackage{tikz}
\usepackage{amsmath,amsfonts,amssymb,amsthm,euscript,braket,xcolor}
\newcommand{\be}{\begin{equation}}
\newcommand{\ee}{\end{equation}}

\newcommand{\Rmnum}[1]{\expandafter\@slowromancap\romannumeral #1@}
\newcommand{\bea}{\begin{eqnarray}}
\newcommand{\eea}{\end{eqnarray}}

\numberwithin{equation}{section}

\usepackage{subfigure}
\begin{document}

\title{\bf Rotating hairy black holes and thermodynamics from gravitational decoupling }

\author{\textbf{Subhash Mahapatra}\thanks{mahapatrasub@nitrkl.ac.in}, \textbf{Indrani Banerjee}\thanks{banerjeein@nitrkl.ac.in}
 \\\\
 \textit{{\small Department of Physics and Astronomy, National Institute of Technology Rourkela, Rourkela - 769008, India}}
}
\date{}


\maketitle
\abstract{ We study the method of extended gravitational decoupling in obtaining static black hole solutions satisfying Einstein's equations with a tensor vacuum. The source has quite generic characteristics and satisfies the strong energy condition. The stationary, axisymmetric counterpart of the static metric is obtained by applying the Newman-Janis and Azreg-A\"{i}nou algorithms. The thermodynamics of the rotating solution is studied and the expressions of various thermodynamic quantities are derived. The dependence of the temperature, free energy and specific heat on the horizon radius is studied for various values of the hairy parameter and the black hole spin. Such a study reveals that small hairy black holes are thermodynamically more stable compared to large hairy black holes, and that the horizon radius and temperature range for which the rotating hairy black holes can be in thermodynamic equilibrium with the surroundings depends non-trivially on the hairy parameters. We further discuss the first law of black hole thermodynamics for the hairy case and discuss its implications.
}

\section{Introduction}
Black holes, one of the most intriguing predictions of general relativity, have received ample observational confirmations over the past five decades. This includes indirect observations in the electromagnetic domain from satellites like RXTE, Chandra, NuStar, etc as well as direct detections like gravitational waves from the LIGO/VIRGO collaboration \cite{LIGOScientific:2016aoc,LIGOScientific:2018mvr,LIGOScientific:2017vwq} and the
black hole image by the Event Horizon Telescope (EHT) \cite{Akiyama:2019cqa,Akiyama:2019brx,Akiyama:2019sww,Akiyama:2019bqs,Akiyama:2019fyp,Akiyama:2019eap}. This proves that black holes are not only  objects of theoretical interest but also have observational relevance. The celebrated no-hair theorem \cite{Bekenstein:1971hc,Israel:1967wq,Israel:1967za,Wald:1971iw,Carter:1971zc,Robinson:1975bv,Mazur:1982db,Mazur:1984wz,Teitelboim:1972qx} propounds that black holes should not carry any charge other than their mass, angular momentum and electric charge. However, the no-hair theorem has not been proved in the rigorous mathematical sense and there exist several counterexamples where the no-hair theorem is violated, see \cite{Volkov:1990sva,Bizon:1990sr,Kuenzle:1990is,Straumann:1990as,Zhou:1991nu,Bizon:1991hw,Bizon:1991nt,Volkov:1995np,Brodbeck:1994vu,Garfinkle:1990qj,Greene:1992fw,Lavrelashvili:1992ia,Torii:1993vm,Mahapatra:2020wym,Erices:2021uyu,Khodadi:2020jij,Khodadi:2021gbc,
Rahmani:2020vvv,
Herdeiro:2015waa,Cunha:2019dwb,Herdeiro:2020wei,Guo:2021zed,
Herdeiro:2014goa,Berti:2013gfa} for a necessarily incomplete selection. In fact, black holes with scalar hairs have been extensively explored as scalar fields play a pivotal role in particle physics and early universe cosmology \cite{Svrcek:2006yi,Peebles:1998qn} and are often invoked to explain the dark sector \cite{McDonald:1993ex,Linde:1982uu}. Further, black holes can harbour charges associated with inner gauge symmetries and they are also known to possess soft quantum hair \cite{Hawking:2016msc}.
With the recent release of black hole images and the discovery of gravitational waves, the scope to test the no-hair theorem has been further enhanced \cite{Herdeiro:2014goa,Berti:2013gfa}.

Apart from introducing various fields, one can evade the no-hair theorem by other means as well. This includes exploring the method of minimal \cite{Ovalle:2017fgl,Ovalle:2007bn} and extended gravitational decoupling \cite{Ovalle:2007bn,Ovalle:2018umz,Ovalle:2018gic,Ovalle:2020kpd,Contreras:2021yxe}. In this procedure, one considers a known solution of Einstein's equations with a given source and then considers including the additional source that in turn modifies the metric. The Schwarzschild metric is generally considered to be the known background. In the case of
 minimal geometric deformation, one modifies only the $g_{rr}$ component of the metric \cite{Ovalle:2017fgl,Casadio:2012pu,Casadio:2012rf,Ovalle:2013vna,Casadio:2013uma,daRocha:2017lqj,Fernandes-Silva:2017nec,Casadio:2017sze,Fernandes-Silva:2018abr,Contreras:2018vph,Contreras:2018gzd,Contreras:2018nfg,
Panotopoulos:2018law,DaRocha:2019fjr,Rincon:2019jal,daRocha:2020rda,Arias:2020hwz,daRocha:2020jdj,daRocha:2020gee,Gabbanelli:2018bhs,Contreras:2019iwm,Abellan:2020jjl,Tello-Ortiz:2020nuc,Zhang:2022niv}, whereas in the extended gravitational decoupling both the $g_{rr}$ and the $g_{tt}$ components of the metric are modified from the Schwarzschild solution. The modification is done in such a way that the additional source satisfies Einstein's equations and the conservation equation with the deformed metric. In this method, Einstein's equations with the additional source completely decouple from Einstein's equations in the vacuum which is solved to obtain the deformed metric. In order to obtain the deformed metric, apart from the Einstein's equations the additional source needs to satisfy certain other conditions: (i) the strong energy condition, (ii) should well behave everywhere outside the horizon and not lead to any additional singularity, and (iii) the corresponding hairy metric should possess a well-defined event horizon.

With these conditions,  novel static, spherically symmetric and asymptotically flat metrics were derived by the method of extended gravitational decoupling in \cite{Ovalle:2020kpd}. In particular, the obtained hairy solutions were not only regular with well-defined event horizons but also satisfied physically relevant and important energy conditions everywhere outside the event horizon. These solutions were constructed in terms of two new parameters $\{\alpha,\ell\}$ (associated with the hair), which could greatly modify the properties of the black hole. For instance, the static hairy solution can become extremal (as opposed to the Schwarzschild case), thereby eminently changing the thermal properties of the black hole. Similarly, the combination $\ell_{0}=\alpha\ell$ quantifies the increase in the black hole entropy from its Schwarzschild value. These observations indicate substantial differences in the thermodynamic as well as structural properties of the hairy black holes compared to the Schwarzschild case. The hairy black hole solutions thus obtained can also have interesting theoretical implications and observational consequences \cite{Islam:2021dyk,Ovalle:2020kpd,Contreras:2021yxe}. Let us also mention that, progress also has been made in recent years to put constraints on the hairy parameters from experimental observations.  The EHT observations related to shadows have been used to set the lower limits for $\ell$ for a given choice of $\alpha$. For instance, for $\alpha=1.0$, the lower limits are $\ell>0.35M$ ($1\sigma$) and $\ell>0.15M$ ($2\sigma$) \cite{Afrin:2021imp,Vagnozzi:2022moj}.

Since astrophysical black holes are rotating in nature it is more realistic and relevant to consider the stationary, axisymmetric counterpart of the aforesaid deformed metric. This is generally accomplished by applying the Newman-Janis algorithm \cite{Newman:1965tw} to the static, spherically symmetric seed metric. The Newman-Janis algorithm has some ambiguities associated with the procedure of complex coordinate transformation \cite{Drake:1998gf} which is circumvented by the method suggested by Azreg-A\"{i}nou which demands that the final rotating metric must be written in the Boyer-Lindquist form and satisfy the gravitational field equations \cite{Azreg-Ainou:2014nra,Azreg-Ainou:2014aqa}. The stationary, axisymmetric solution can also be obtained by the method of gravitational decoupling which has been discussed in \cite{Contreras:2021yxe}. Here the axisymmetric metric assumes the Gurses-Gursey form with a general mass function $\Tilde{m}(r)$. Interestingly, the Einstein tensor computed with this metric depends linearly on the derivatives of $\Tilde{m}(r)$. This enables one to split $\Tilde{m}(r)$ as $\Tilde{m}(r)=m(r) + \alpha m_s(r)$ where $m(r)$ is associated with the axisymmetric seed metric while $m_s(r)$ represents the deformation in the mass function due to the presence of the additional energy momentum tensor.

In this work, we perform this analysis and obtain the stationary, axisymmetric counterpart of the gravitational decoupling inspired hairy black holes by applying the Azreg-A\"{i}nou formalism and find analytic expressions of rotating hairy black holes satisfying various energy conditions outside the event horizon.  We provide a detailed discussion on the rotating solution for the case in which the hair satisfies the strong energy condition, whereas a brief but analogous and straightforward discussion for the case in which the hair satisfies the dominant energy condition is presented in the Appendix. The obtained solution makes sure that the source satisfies the energy conditions (strong or dominant) everywhere outside the horizon. This is important considering that it is always physically desirable to have the obtained gravity solution with additional matter fields respecting the relevant energy conditions. Moreover, we also find the form of complexified null tetrads which leads to the same rotating hairy solution, however now, from the standard Newman-Janis procedure.

Once the stationary, axisymmetric counterpart of the seed metric is obtained, we study the horizon structure and the extremality conditions associated with the rotating black hole. We next explore the curvature invariants like the Ricci scalar and the Kretschmann scalar which are finite everywhere except at the origin at $r=0$. Finally, we investigate the thermodynamics of the rotating black hole solution thus derived. We Euclideanize the metric and present the expressions for entropy, temperature and angular velocity. We further find the expression of quasilocal energy in the slow rotating approximation and find corrections due to hair.  At the horizon, the quasilocal energy is found to be (twice) the irreducible mass, whereas asymptotically it reduces to the Arnowitt-Deser-Misner (ADM) mass. The variation of the Hawking temperature with the horizon radius is presented for different values of spin $a$ and parameters $\{\alpha,\ell\}$. This enables us to comment on the stability of the black holes which can be further established when we analyze the free energy and specific heat of the system. Our analysis reveals that small hairy black holes for which the temperature increases with the horizon radius are more stable compared to large hairy black holes for which the temperature decreases with the horizon radius. This result is further confirmed by plotting the free energy and specific heat with the horizon radius for various values of $\alpha$, $\ell$ and $J$. The specific heat is positive for small black holes and negative for the larger ones thereby establishing that small black holes are indeed thermodynamically more stable. We further analyse the free energy in the fixed momentum canonical ensemble and find that small black holes always have smaller free energy compared to large black holes at a fixed temperature. Our study further suggests that the horizon radius and temperature range for the stable rotating hairy black holes depend non-trivially on the hairy parameters. Finally, we also present the first law of black hole thermodynamics for the hairy case both for rotating and non-rotating black holes.

The paper is organized in the following way: In Section \ref{S2} we present the static, spherically symmetric hairy black hole solutions obtained by the method of extended gravitational decoupling on the seed Schwarzschild metric. The stationary, axisymmetric counterparts of the static black hole solutions are derived using the Newman-Janis method/Azreg-A\"{i}nou approach in Section \ref{S3}. Section \ref{S4} is dedicated to investigating the thermodynamics of the aforesaid black hole solution which enables us to understand the thermodynamic stability of these black holes. Finally, we conclude with a summary of our results with some scope for future work in \ref{S5}.

\section{A brief review of the static hairy solutions}
\label{S2}
In this section, we review the hairy black hole solutions obtained in \cite{Ovalle:2020kpd} using the extended gravitational decoupling (EGD) technique. The essential idea of the EGD technique is to consider the seed solutions of the Einstein field equations and make a consistent transformation of the metric functions such that the total energy-momentum tensor separates out into subparts that are then computationally more tractable. To see how this works, let us start with Einstein field equation with source $\tau_{\mu\nu}$,
\begin{eqnarray}
G_{\mu\nu}=R_{\mu\nu}-\frac{1}{2}g_{\mu\nu} R = \kappa^2 {\tau}_{\mu\nu} \,,
\label{EinsteinEOM}
\end{eqnarray}
where $\kappa^2=8\pi G_N$, and $G_N$ is the Newton's constant. We consider the situation where the static, spherically symmetric solution of (\ref{EinsteinEOM}) has been worked out and is given by,
\begin{eqnarray}
ds^2=e^{g_0(r)}dt^2 -e^{f_0(r)}dr^2 - r^2\left(d\theta^2 + \sin^2\theta d\phi^2  \right)\,,
\label{metric1}
\end{eqnarray}
where the form of $g_0(r)$ and $f_0(r)$ are known and the above metric is termed as the seed metric.

We next consider adding new fields or hair in the system such that the total energy momentum tensor becomes,
\begin{align}
\tilde{\tau}_{\mu\nu}=\tau_{\mu\nu} + \chi_{\mu\nu} \,,
\label{1}
\end{align}

When the source term $\chi_{\mu\nu}$ associated with the hair is added, the new static and spherically symmetric solution of Einstein's equations assumes the form,
\begin{eqnarray}
ds^2=e^{g(r)}dt^2 -e^{f(r)}dr^2 - r^2\left(d\theta^2 + \sin^2\theta d\phi^2  \right)\,.
\label{metric2}
\end{eqnarray}
The above metric is obtained by extended geometric deformation of the seed metric such that,
\begin{align}
g(r)=g_0(r)+\alpha \mathcal {G}(r) \,, \nonumber \\
e^{-f(r)}=e^{-f_0(r)} +  \alpha \mathcal{F}(r) \,,
\label{3}
\end{align}
where $\mathcal{G}(r)$ and $\mathcal{F}(r)$ are respectively associated with the geometric deformation of the $g_{tt}$ and $g_{rr}$ components of the seed metric while $\alpha$ is a parameter which keeps track of the deformation. Now, for a general static and spherically symmetric metric of the form (\ref{metric2}), the Einstein equation reduces to
\begin{eqnarray}
& & \kappa^2 \left( \tau_{0}^{\ 0} + \chi_{0}^{\ 0} \right) = \frac{1}{r^2} - e^{-f(r)} \left(\frac{1}{r^2} -\frac{f'(r)}{r}  \right) \,, \nonumber \\
& & \kappa^2 \left( \tau_{1}^{\ 1} + \chi_{1}^{\ 1} \right) = \frac{1}{r^2} - e^{-f(r)} \left(\frac{1}{r^2} + \frac{g'(r)}{r}  \right) \,, \nonumber \\
& & \kappa^2 \left( \tau_{2}^{\ 2} + \chi_{2}^{\ 2} \right) =  - \frac{e^{-f(r)}}{4} \left(2g''(r) + g'^2(r)-f'(r)g'(r)+2\frac{g'(r)-f'(r)}{r}  \right) \,,
\label{EinsteinEOM1}
\end{eqnarray}
where the prime denotes the derivative with respect to $r$, and $\tilde{\tau}_{3}^{\ 3}=\tilde{\tau}_{2}^{\ 2}$ due to the spherical symmetry.

The transformation in Eq.~(\ref{3}) ensures that the two sources $\tau_{\mu\nu}$ and $\chi_{\mu\nu}$ are decoupled completely such that
the set of equations containing the source $\chi_{\mu\nu}$ are given by,
\begin{eqnarray}
& & \kappa^2 \chi_{0}^{\ 0} =-\alpha \left( \frac{\mathcal{F}(r)}{r^2}+\frac{\mathcal{F}'(r)}{r} \right) \,, \nonumber \\
& & \kappa^2 \chi_{1}^{\ 1} + \alpha Z_1 = -\alpha \mathcal{F}(r) \left(\frac{1}{r^2}+\frac{g'(r)}{r^2} \right) \,, \nonumber \\
& & \kappa^2 \chi_{2}^{\ 2}  + \alpha Z_2 = -\alpha \frac{\mathcal{F}(r)}{4}\left(2 g''(r) + g'^2(r) + \frac{2g'(r)}{r} \right) -\alpha \frac{\mathcal{F}'(r)}{4}\left(g'(r) + \frac{2}{r}\right)\,, \nonumber \\
\label{EinsteinEOMsource1}
\end{eqnarray}
where
\begin{eqnarray}
& & Z_1 = \frac{e^{-f_0(r)}\mathcal{G}'(r)}{r} \,, \nonumber \\
& & Z_2 =e^{-f_0(r)} \left( 2\mathcal{G}''(r) + \mathcal{G}'^2(r) + \frac{2 \mathcal{G}'(r)}{r} + 2 g_{0}'(r) \mathcal{G}'(r) - f_{0}'(r)\mathcal{G}'(r) \right)  \,.
\label{EinsteinEOMsource2}
\end{eqnarray}
Therefore, it might be possible to compute the deformed hairy metric solution if the deformation functions $\mathcal{G}$ and $\mathcal{F}$ can be computed from the tensor $\chi_{\mu\nu}$ using Eq.~(\ref{EinsteinEOMsource1}).

The conservation of the total energy momentum tensor can be written as,
\begin{align}
&\Tilde{\nabla}_\sigma \tilde{\tau}^{\sigma}_{\alpha}=0 \,,\nonumber \\
&\Tilde{\nabla}_\sigma (\tau^{\sigma}_{\alpha} + \chi^{\sigma}_{\alpha})=0\,,
\end{align}
where the covariant derivative $\Tilde{\nabla}$ is computed with the deformed metric (\ref{metric2}).
Note that the energy momentum tensor $\tau^{\sigma\alpha}$ is separately conserved
\begin{align}
&\nabla_\sigma \tau^{\sigma}_{\alpha} =0\,,
\end{align}
the covariant derivative being calculated with the seed metric (\ref{metric1}).

In case of extended gravitational decoupling $\tau^{\sigma\alpha}$ is not conserved with respect to the deformed metric and is given by,
\begin{align}
\Tilde{\nabla}_\sigma \tau^{\sigma}_{\alpha} = {\nabla}_\sigma \tau^{\sigma}_{\alpha} -\alpha \frac{\mathcal{G}^\prime(r)}{2}(\tau^0_0 -\tau^1_1) \delta^1_\alpha =\alpha \frac{\mathcal{G}^\prime(r)}{2}(\tau^0_0 -\tau^1_1) \delta^1_\alpha \,.
\label{Eq6}
\end{align}
It is important to note that for minimal gravitational decoupling $\mathcal{G}(r)=0$ such that the energy momentum tensor $\tau^\mu_\nu$ is also conserved with respect to the deformed metric as well (see (\ref{Eq6})).

Applying the above procedure for the simplest case of interest, namely $\tau_{\mu\nu}=0$ and $\chi_{\mu\nu}\neq 0$, gives us the Schwarzschild solution as the seed metric having,
\begin{eqnarray}
& & e^{g_0(r)}=e^{-f_0(r)} = 1 - \frac{2 M}{r} \,.
\label{seedsol}
\end{eqnarray}
The conditions $\tau_{\mu\nu}=0$ and $\chi_{\mu\nu}\neq 0$ correspond to the physical situation where the Schwarzschild vacuum is filled with a tensor vacuum $\chi_{\mu\nu}$. The corresponding hairy solution can be obtained by solving Eq.~(\ref{EinsteinEOMsource1}) for the two deformation functions $\mathcal{G}$ and $\mathcal{F}$ for the given tensor vacuum $\chi_{\mu\nu}$.

For simplicity and to proceed further, one assumes that the deformed metric satisfies the condition:
\begin{align}
e^{g(r)}=e^{-f(r)}\,.
\label{Eq1}
\end{align}
The above condition when applied in Eq.~(\ref{EinsteinEOM1}) immediately gives rise to the following equation of state:
\begin{align}
\tilde{p}_r=-\tilde{\rho}\,.
\label{Eq2}
\end{align}
Using Eqs.~(\ref{Eq1}) and (\ref{Eq2}) one obtains the following relation between the deformation functions,
\begin{eqnarray}
& & \alpha \mathcal{F}(r) = \left(1-\frac{2M}{r}\right)\left(e^{\alpha\mathcal{G}(r)} -1 \right)\,,
\label{Eq2-1}
\end{eqnarray}
such that the line element of the deformed metric (\ref{metric2}) becomes
\begin{eqnarray}
ds^2=\left(1-\frac{2M}{r}\right)e^{\alpha\mathcal{G}(r)} dt^2 - \left(1-\frac{2M}{r}\right)^{-1} e^{-\alpha\mathcal{G}(r)} dr^2 - r^2\left(d\theta^2 + \sin^2\theta d\phi^2  \right)\,.
\label{seedmetric}
\end{eqnarray}
Therefore, one is required to find the deformation function $\mathcal{G}$ and the three components of $\chi_{\mu\nu}$, which must satisfy the Eq.~(\ref{EinsteinEOMsource1}). To proceed further, we need to impose physically motivated constraints on $\mathcal{G}(r)$ and $\chi_{\mu\nu}$. As in \cite{Ovalle:2020kpd}, we now discuss several solutions of (\ref{seedmetric}) based on different imposed conditions.

\subsection{Kiselev black hole}
The gravitational decoupling procedure can also be used to generate the Kiselev black hole solution. For instance, let us apply the following condition on the perturbed energy momentum tensor:
\begin{align}
\chi^0_0=a\chi^1_1 + b \chi^2_2 \,,
\end{align}
where $a$ and $b$ are constants. Using the above condition in (\ref{EinsteinEOMsource1}) one obtains the following form of the deformed metric given in (\ref{seedmetric}),
\begin{align}
ds^2=\bigg(1-\frac{2\tilde{M}}{r} + \frac{l^n}{r^n}\bigg)dt^2  - \bigg(1-\frac{2\tilde{M}}{r} + \frac{l^n}{r^n}\bigg)^{-1}dr^2 - r^2 d\Omega^2 \,,
\label{Kiselev}
\end{align}
where $\tilde{M}=M+\frac{l_0}{2}$, $n=\frac{2}{b}(a-1)$, while $l_0$ and $l$ are constants  proportional to $\alpha$. Imposing the dominant energy condition on the energy momentum tensor and invoking asymptotic flatness demand that $1\leq n \leq 2$. When $n=1$, one retrieves the Schwarzschild solution while $n=2$ yields a metric similar to the Reissner Nordstr\"{o}m solution with traceless $\chi_{\mu\nu}$.
The horizon radius is obtained by solving for $g^{rr}(r=r_h)=0$, which in the present case requires one to solve for $r_h$ in the equation,
\begin{align}
r_{h}^n-2 \tilde{M}r_{h}^{n-1}+l^n=0 \,.
\end{align}

\subsection{Hairy black hole with strong energy condition}
\item Another route to obtain solution for (\ref{Eq2-1}) is given by imposing strong energy condition on $\chi_{\mu\nu}$ in the region of spacetime accessible to an observer outside the horizon,
\begin{eqnarray}
& & \tilde{\rho} + \tilde{p}_r + 2\tilde{p}_\theta \geq 0, ~~~  \tilde{\rho} + \tilde{p}_r  \geq 0, ~~~ \tilde{\rho} + \tilde{p}_\theta \geq 0 \,.
\label{SEC}
\end{eqnarray}
Then the conditions in (\ref{Eq1}) and (\ref{Eq2}) imply,
\begin{eqnarray}
& & \chi_{2}^{\ 2} \leq 0, ~~~  \chi_{0}^{\ 0} \geq  \chi_{2}^{\ 2} \,.
\label{thetaconstraintSEC}
\end{eqnarray}
Together with Eq.~(\ref{EinsteinEOMsource1}), the above two constraint conditions can be reformulated as,
\begin{eqnarray}
& & S_1(r) \equiv (r-2M)h''(r) + 2h'(r) \geq 0 \,,  \nonumber \\
& & S_2(r) \equiv r(r-2M)h''(r) + 4 M h'(r)-2h(r)+2 \geq 0 \,,
\label{S1S2eqom}
\end{eqnarray}
where $h(r)=e^{\alpha \mathcal{G}(r)}$ has been used. Demanding the black hole solution with a proper horizon at $r\sim 2M$ and that the spacetime solution approaches the standard
Schwarzschild solution in the asymptotically flat $r\gg 2M $ regime, the following form of the deformed metric is obtained,
\begin{eqnarray}
ds^2=\bigg(1-\frac{2\mathcal{M}}{r} + \alpha e^{\frac{-r}{\mathcal{M} - \alpha \ell/2}}\bigg)dt^2 - \bigg(1-\frac{2\mathcal{M}}{r} + \alpha e^{\frac{-r}{\mathcal{M} - \alpha \ell/2}}\bigg)^{-1}dr^2 - r^2 d\Omega^2 \,,
\label{metfunclSEC}
\end{eqnarray}
where the shift mass term $\mathcal{M}=M+\alpha \ell/2$, due to black hole hair is introduced. This is the mass that will be observed by an asymptotic observer at infinity. The effective density and tangential pressure are respectively given by,
\begin{eqnarray}
& & \tilde{\rho} = - \tilde{p}_r = \chi_{0}^{\ 0}=\frac{\alpha e^{-r/M}}{\kappa^2 M r^2} (r-M) \,,  \nonumber \\
& &  \tilde{p}_\theta = -\chi_{2}^{\ 2} = \frac{\alpha e^{-r/M}}{2\kappa^2 M r^2} (r-2M) \,.
\label{densitypressureSEC}
\end{eqnarray}
The horizon radius can be obtained by solving for $g^{rr}=0$ which leads to,
\begin{align}
r_h-2\mathcal{M}+\alpha r_h e^{-r_h/M}=0\,.
\label{Eq3}
\end{align}
For the horizon radius $r_h\geq 2M$, the hairy solution satisfies the strong energy condition. Moreover, consideration of the extremal case, i.e., $r_h=2M$ in (\ref{Eq3}) provides a lower bound on $\ell$ which is given by,
\begin{align}
\ell \geq \frac{2M}{e^2}\,.
\label{Eq4}
\end{align}

From the above discussion, it is therefore clear that the metric solution obtained in Eq.~(\ref{metfunclSEC}) corresponds to a consistent hairy black hole solution with parameter $\{M, \alpha, \ell\}$. The parameter $\ell$ denote the charge that can be associated with the primary hair. With the hairy black holes in hand, we can now try to find the axisymmetric counterpart of this solution. In the process, we will also generate the axisymmetric counterpart of the Kiselev black hole (\ref{Kiselev}).

\subsection{Generating rotating solutions from the static metric}
\label{S3}
In this section, we generate stationary, axisymmetric solutions of the gravitational field equations from a given static, spherically symmetric solution of the same. The Newman-Janis algorithm is generally used for this purpose. However, there is an ambiguity in the algorithm regarding the complexification procedure which has been discussed and addressed in \cite{Drake:1998gf}.
In this paper, we follow the method suggested by \cite{Azreg-Ainou:2014nra} to generate the rotating metric.
We consider the following static metrics discussed in the last section as the seed metric on which we apply the complex coordinate transformation,
\begin{align}
ds^2=\bigg(1-\frac{2\tilde{M}}{r} +\frac{l^n}{r^n}\bigg)dt^2 - \frac{dr^2}{\bigg(1-\frac{2\tilde{M}}{r} +\frac{l^n}{r^n}\bigg)}
- r^2 d\Omega^2 \,,
\label{3-1}
\end{align}
\begin{align}
ds^2=\bigg(1-\frac{2\mathcal{M}}{r}+ \alpha e^{-r/(\mathcal{M}-\alpha l/2)}\bigg)dt^2 - \frac{dr^2}{\bigg(1-\frac{2\mathcal{M}}{r}+ \alpha e^{-r/(\mathcal{M}-\alpha l/2)}\bigg)} - r^2 d\Omega^2 \,.
\label{3-2}
\end{align}
To see how the algorithm works we consider a general static metric,
\begin{align}
ds^2=e^{2\nu(r)}dt^2 - e^{2\lambda(r)}dr^2 -\gamma(r)(d\theta^2 + sin^2\theta d\phi^2)\,.
\label{3-4}
\end{align}
We next write the metric in advanced null coordinates,
\begin{align}
ds^2=e^{2\nu(r)}du^2 + 2 e^{\nu(r)+\lambda(r)}dudr -\gamma(r)(d\theta^2 + sin^2\theta d\phi^2)\,.
\label{3-5}
\end{align}
The contravariant components of the metric in (\ref{3-5}) can be written in terms of null tetrads,
\begin{align}
g^{\mu\nu}=l^\mu n^\nu + l^\nu n^\mu -m^\mu \bar{m}^\nu -m^\nu \bar{m}^\mu \,,
\label{3-6}
\end{align}
where,
\begin{align}
l^\mu&=\delta^\mu_r \,, \\
n^\mu&=e^{-\lambda-\nu}\delta^\mu_u -\frac{1}{2}e^{-2\lambda}\delta^\mu_r \,, \\
m^\mu&=\frac{1}{\sqrt{2\gamma(r)}}\bigg(\delta^\mu_\theta  + \frac{i}{sin\theta}\delta^\mu_\phi\bigg) \,,
\label{3-7}
\end{align}
such that,
\begin{align}
l^\mu l_\mu=n^\mu n_\mu=m^\mu m_\mu=0,~~~~~l^\mu n_\mu =-m^\mu \bar{m}_\mu=1,~~~~l^\mu m_\mu=n^\mu m_\mu =0 \,.
\label{3-8}
\end{align}
Let us consider now a complex coordinate transformation,
\begin{align}
x'^\mu=x^\mu + i y^\mu (x^\nu) \,.
\label{3-9}
\end{align}
A special case of the above transformation is known as the Newman-Janis transformation, which is given by,
\begin{align}
r^\prime=r + i a cos\theta; ~~~u^\prime =u-iacos\theta \,.
\label{3-10}
\end{align}
Denoting the null tetrads by the shorthand,
\begin{align}
V^\mu_a\equiv (l^\mu,n^\mu,m^\mu,\bar{m}^\mu) \,,
\label{3-11}
\end{align}
we then transform the null tetrads according to Newman-Janis transformation, such that,
\begin{align}
V^{\prime\mu}_a=\frac{\partial x^{\prime \mu}}{\partial x^\nu}V^\nu_a \,.
\label{3-12}
\end{align}
After the Newman-Janis transformation the null tetrads assume the form,
\begin{align}
l^{\prime\mu}&=\delta^\mu_r \,, \nonumber \\
n^{\prime\mu}&=\sqrt{\frac{\Upsilon(a,r,\theta)}{\Gamma(a,r,\theta)}}\delta^\mu_u-\frac{\Upsilon(a,r,\theta)}{2}\delta^\mu_r \,, \nonumber \\
m^{\prime\mu}&=\frac{1}{\sqrt{2\Phi(a,r,\theta)}}\bigg[\delta^\mu_\theta + i a \sin\theta(\delta^\mu_u -\delta^\mu_r)+\frac{i}{\sin\theta}\delta^\mu_\phi \bigg] \,,
\label{3-13}
\end{align}
such that there are the following restrictions on $\Upsilon$, $\Gamma$ and $\Phi$,
\begin{align}
\lim_{a\to 0} \Upsilon(a,r,\theta)=e^{-2\lambda(r)},~~~ \lim_{a\to 0}\Gamma (a,r,\theta)=e^{2\nu(r)}, ~~~\lim_{a\to 0}\Phi (a,r,\theta)= \gamma(r) \,.
\label{3-14}
\end{align}
The original Newman-Janis approach has a certain level of ambiguity associated with the complexification of $r$, which in turn determines the forms of $\Upsilon$, $\Gamma$ and $\Phi$. The complexification procedure is completely arbitrary and requires clever guess work. Consequently, the resultant axisymmetric solution thus obtained often cannot be cast to the Boyer-Lindquist form or fail to satisfy the original field equations. This is because {$\Gamma(r, \theta, a),\Upsilon(r, \theta, a), \Phi(r, \theta, a)$} are fixed by the complexification procedure and there remain no free parameters or functions
to act on to achieve the transformation to Boyer-Lindquist coordinates.

Since it is difficult to guess the right complexification procedure before hand, Azreg-A\"{i}nou proposed  to keep $\Upsilon$, $\Gamma$ and $\Phi$ as general functions of $r$ and $\theta$. This should be contrasted with the original Newman-Janis algorithm where $\Upsilon$, $\Gamma$ and $\Phi$ respectively are derived from $e^{-2\lambda(r)}$, $e^{2\nu(r)}$ and $\gamma(r)$. Demanding that the metric must be cast in the Boyer-Lindquist form which fixes $\Upsilon$ and $\Gamma$ in terms of the components of the static seed metric. The form of $\Phi$ is fixed by demanding that the axisymmetric metric thus obtained is a solution of the original gravitational field equations. Below we explicitly show how the algorithm proposed by Azreg-A\"{i}nou works in generating stationary axisymmetric solutions from a static seed metric.


Using (\ref{3-13}) we can construct the stationary, axisymmetric form of the metric in advanced null coordinates $(u,r,\theta,\phi)$,
\begin{align}
g^{uu}(r,\theta)&=-\frac{a^2 \sin^2\theta}{\Phi}\,,~~~~~~g^{ur}(r,\theta)=\sqrt{\frac{\Upsilon}{\Gamma}}+\frac{a^2 \sin\theta}{\Phi}\,,~~~~~~g^{u\phi}(r,\theta)=-\frac{a}{\Phi} \,, \nonumber\\
g^{rr}(r,\theta)&=-\Upsilon-\frac{a^2 \sin^2\theta}{\Phi}\,,~~~~~~g^{r\phi}(r,\theta)=\frac{a}{\Phi}\,,\nonumber\\
g^{\theta\theta}(r,\theta)&=-\frac{1}{\Phi}\,,~~~~~~g^{\phi\phi}(r,\theta)=-\frac{1}{\Phi \sin^2\theta} \,.
\label{3-15}
\end{align}
We invert (\ref{3-15}) to obtain the covariant components of the metric in Eddington Finkelstein coordinates,
\begin{align}
g_{uu}(r,\theta)&=\Gamma\,,~~~~~~g_{ur}(r,\theta)=\sqrt{\frac{\Gamma}{\Upsilon}}\,,~~~~~~g_{u\phi}(r,\theta)=a \sin^2\theta\bigg(\sqrt{\frac{\Gamma}{\Upsilon}}-\Gamma\bigg)\,, \nonumber \\
g_{r\phi}(r,\theta)&=-a \sin^2\theta\sqrt{\frac{\Gamma}{\Upsilon}}\,,~~g_{\theta\theta}(r,\theta)=-\Phi\,,~~g_{\phi\phi}(r,\theta)=-\sin^2\theta\biggl[ \Phi + a^2\sin^2\theta\bigg(2\sqrt{\frac{\Gamma}{\Upsilon}}-\Gamma\bigg)  \biggr]\,.
\label{3-16}
\end{align}
In order to write the rotating solution in the Boyer-Lindquist form, we consider the coordinate transformation where,
\begin{align}
du=dt-\frac{(K + a^2)dr}{e^{-2\lambda}\gamma + a^2};~~~~~~d\phi=d\varphi-\frac{a dr}{ e^{-2\lambda}\gamma+a^2} \,,
\label{3-17}
\end{align}
where,
\begin{align}
K(r)=\gamma(r)e^{-\nu(r)-\lambda(r)} \,.
\label{3-17-1}
\end{align}
In the Boyer-Lindquist coordinates, the axisymmetric metric has only one off-diagonal term i.e, $g_{t\phi}$. Therefore, after applying the coordinate transformation (\ref{3-17}), we put the other off-diagonal components to zero. From this we obtain the expression for $\Upsilon(r,\theta)$ and $\Gamma(r,\theta)$ as,
\begin{eqnarray}
& & \Gamma(r,\theta)=\frac{(e^{-2\lambda(r)} \gamma(r)+a^2 \cos^2\theta)\Phi(r,\theta)}{(K(r)+a^2cos^2\theta)^2} \,,\nonumber\\
&  & \Upsilon(r,\theta)=\frac{e^{-2\lambda(r)} \gamma(r)+a^2cos^2\theta}{\Phi(r,\theta)} \,.
\label{3-18}
\end{eqnarray}
Using (\ref{3-17}) and (\ref{3-18}), we obtain the final form of the stationary, axisymmetric metric in Boyer-Lindquist coordinates,
\begin{align}
ds^2=\Gamma(r,\theta)dt^2  - \frac{\Phi (r,\theta)}{e^{-2\lambda(r)}\gamma(r)+a^2}dr^2 + 2a\sin^2 \theta \bigg[\frac{K(r)-e^{-2\lambda(r)} \gamma(r)}{\left(K(r)+a^2cos^2\theta\right)^2}\bigg]\Phi (r,\theta)dtd\varphi \nonumber \\
 -\Phi(r,\theta) d\theta^2 -\Phi(r,\theta) \sin^2\theta\bigg[1+a^2\sin^2\theta\frac{2 K(r)+a^2\cos^2\theta-e^{-2\lambda(r)}\gamma(r)}{\left(k(r)+a^2cos^2\theta\right)^2}\bigg]d\varphi^2 \,.
\label{3-19}
\end{align}
The only unknown quantity in (\ref{3-19}) is the functional form of $\Phi(r,\theta)$. This is determined from the Einstein field equations with source $\tilde{\tau}_{\mu\nu}$. It is important to note that if we compute the Einstein tensor from (\ref{3-19}), it has the non-zero off-diagonal component $G_{r\theta}$. Since $\tilde{\tau}_{r\theta}$ is zero, therefore $G_{r\theta}$ should vanish to retain consistency with the Einstein field equations. This gives rise to the following differential equation for $\Phi(r,\theta)$,
\begin{align}
\frac{\Phi^2}{\Sigma^2}(-48 a^2 r \sin2\theta \Phi^2 -24\Phi_{,r}\Phi_{,\theta}\Sigma^2+4\Phi_{,r\theta}\Sigma^2)=0 \,,
\label{3-20}
\end{align}
where $\Sigma(r,\theta)=r^2 + a^2 \cos^2\theta$. Solving this equation gives,
\begin{align}
\Phi(r,\theta)=r^2 + a^2 cos^2\theta=\Sigma(r,\theta) \,.
\label{3-21}
\end{align}
Having obtained the form of $\Phi(r,\theta)$, we now substitute the form of the metric components (\ref{3-1}) and (\ref{3-2}) in (\ref{3-17-1}) and (\ref{3-18}) to obtain their corresponding rotating counterparts.\\

First of all we note that for our main cases of interest (Kiselev and hairy black holes), the $g_{tt}$ and $g_{rr}$ components of the static black holes are inverse of each other, i.e. $e^{2\nu}=e^{-2\lambda}$. Thus,
\begin{align}
K(r)=\gamma(r)\,.
\label{3-22}
\end{align}
Moreover, we also have $\gamma(r)=r^2$. We now have all the ingredients to write down the axisymmetric counterpart of static black holes in (\ref{3-1}) and (\ref{3-2}). A general axisymmetric metric has the form,
\begin{align}
ds^2= g_{tt}(r,\theta)dt^2 + 2g_{t\varphi}(r,\theta) dtd\varphi + g_{\varphi\varphi}(r,\theta)d\varphi^2 + g_{rr}(r,\theta) dr^2 +g_{\theta\theta}(r,\theta)d\theta^2 \,.
\label{3-23}
\end{align}
For the Kiselev static black hole metric in (\ref{3-1}), we have
\begin{align}
g_{tt}(r,\theta)&= \frac{r^2 \big(1-2\tilde{M}/r + l^n/r^n\big)+ a^2 \cos^2\theta}{\Sigma} \,, \nonumber \\
g_{t\varphi}(r,\theta)&=\frac{ar^2\sin^2\theta}{\Sigma}\bigg(\frac{2\tilde{M}}{r}-\frac{l^n}{r^n}\bigg) \,,\nonumber \\
g_{\varphi\varphi}(r,\theta)&=-\Sigma \sin^2\theta\bigg[1 + a^2\sin^2\theta\frac{r^2(1+2\tilde{M}/r-l^n/r^n) +a^2\cos^2\theta }{\Sigma^2}\bigg] \,, \nonumber \\
g_{rr}(r,\theta)&=-\frac{\Sigma}{a^2+r^2\big(1-2\tilde{M}/r+l^n/r^n\big)} \,, \nonumber\\
g_{\theta\theta}(r,\theta)&=-\Sigma \,.
\label{rotatingkiselevmet}
\end{align}
Similarly, for the static hairy black hole metric in (\ref{3-2}), we have
\begin{align}
g_{tt}(r,\theta)&= \frac{r^2 \big(1-2\mathcal{M}/r + \alpha e^{-r/(\mathcal{M}-\alpha l/2)}\big)+ a^2 \cos^2\theta}{\Sigma}\,, \nonumber \\
g_{t\varphi}(r,\theta)&=\frac{ar^2 \sin^2\theta}{\Sigma}\bigg(\frac{2\mathcal{M}}{r}-\alpha e^{-r/(\mathcal{M}-\alpha l/2)}\bigg)\,,\nonumber \\
g_{\varphi\varphi}(r,\theta)&=-\Sigma \sin^2\theta\bigg[1 + a^2 \sin^2\theta\frac{r^2(1+2\mathcal{M}/r-\alpha e^{-r/(\mathcal{M}-\alpha l/2)}) +a^2 \cos^2\theta }{\Sigma^2}\bigg]\,,\nonumber \\
g_{rr}(r,\theta)&=-\frac{\Sigma}{a^2+r^2\big(1-2\mathcal{M}/r+ \alpha e^{-r/(\mathcal{M}-\alpha l/2)}\big)}\,, \nonumber\\
g_{\theta\theta}(r,\theta)&=-\Sigma \,.
\label{rotatinghairmet}
\end{align}
Notice that all the metric coefficients, except the $g_{\theta\theta}$ term, receive corrections due to hair. As expected, this new rotating hairy solution reduces to the Kerr geometry for $\alpha=0$, and approaches the Schwarzschild geometry asymptotically with mass $\mathcal{M}$.

The above stationary, axisymmetric solution can also be obtained by the method of gravitational decoupling as done in \cite{Contreras:2021yxe}. In this method the axisymmetric solution assumes the Gurses-Gursey form with a general mass function $\Tilde{m}(r)$. Interestingly, the Einstein tensor evaluated with the aforesaid metric turns out to possess only linear derivatives of $\Tilde{m}(r)$, although it depends non-linearly on the spin parameter $\Tilde{a}$. Thus, the mass function can be split in the form $\Tilde{m}(r)=m(r)+\alpha m_s(r)$, where $m(r)$ represents the mass function of the axisymmetric seed metric while the deformed metric corresponding to the additional energy momentum tensor is obtained by adding $m_s(r)$ to $m(r)$.   The rotating hairy metric so obtained takes the same form as in Eq.~(\ref{rotatinghairmet}). These results for the hairy rotating metric from two seemingly different methods not only emphasize the correctness of the obtained rotating solution but also suggest that these two methods complement each other well to generate rotating solutions in the present context.

At this point, to avoid any confusion, we would like to emphasize the difference between the various mass parameters $\{M,\mathcal{M},\tilde{M}\}$ appearing in Eqs.~(\ref{rotatingkiselevmet}) and (\ref{rotatinghairmet}). The mass parameter $M$ corresponds to the usual Schwarzschild mass, i.e. of the non-hairy black hole. Whereas the parameter $\mathcal{M}=M+\alpha \ell/2$ corresponds to the ADM mass of the hairy black hole, which reduces to the Schwarzschild mass $M$ when the hairy parameter $\alpha$ goes to zero. On the other hand, the parameter $\tilde{M}$ arises in the discussion of the Kiselev black only. Accordingly, it does not appear and play any role in the discussion of the hairy black holes below.

\begin{figure}[h!]
\begin{minipage}[b]{0.5\linewidth}
\centering
\includegraphics[width=2.8in,height=2.3in]{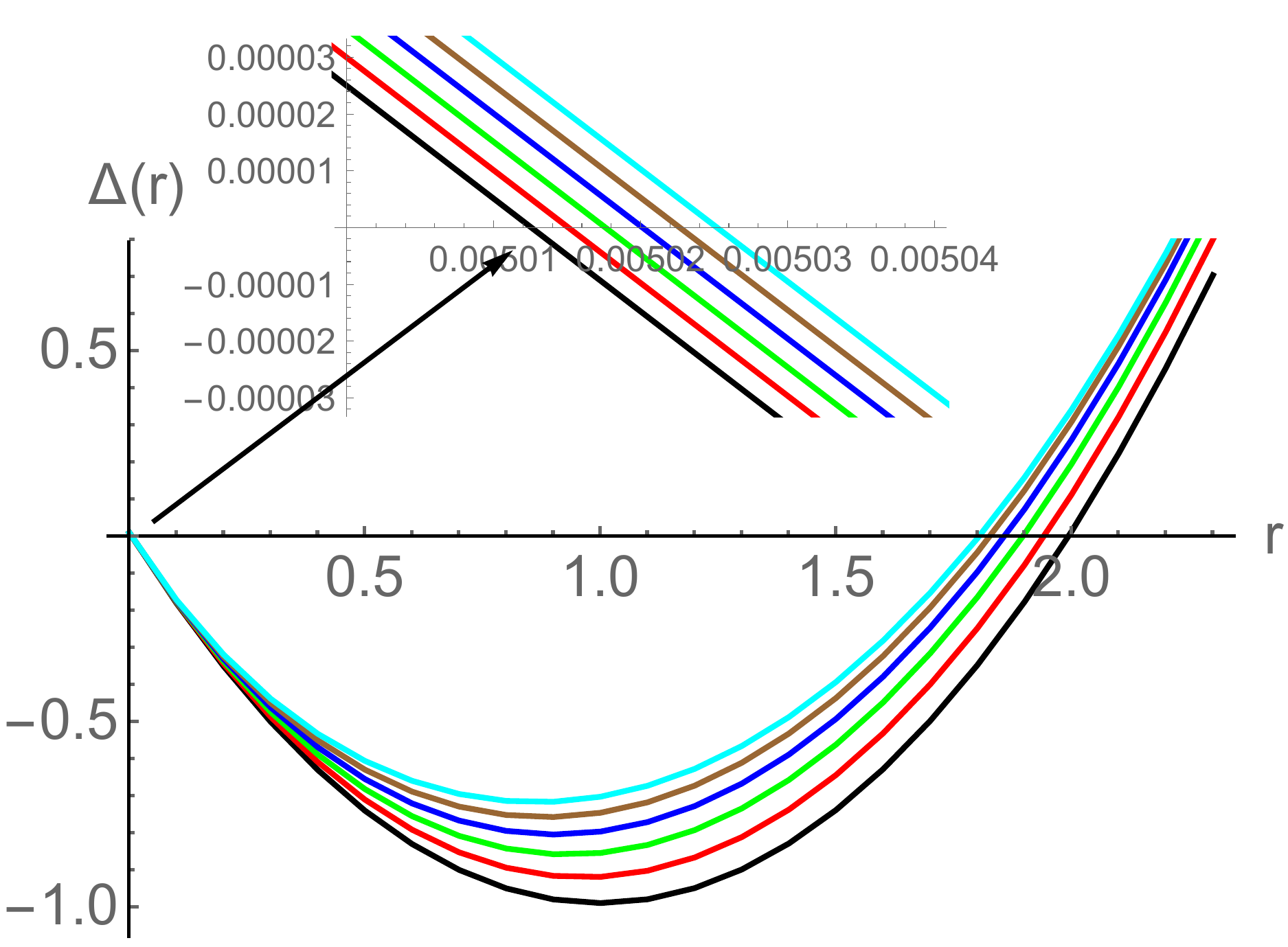}
\caption{ \small The behaviour of $\Delta(r)$ as a function of $r$ for different values of $\alpha$. Here $\mathcal{M}=1$, $\ell=0.4$, $a=0.1$ are used. The black, red, green, blue, brown and cyan curves correspond to $\alpha=0$, $0.2$, $0.4$, $0.6$, $0.8$ and $1.0$ respectively. }
\label{rvsdelvsalphaLPt2aPt1}
\end{minipage}
\hspace{0.4cm}
\begin{minipage}[b]{0.5\linewidth}
\centering
\includegraphics[width=2.8in,height=2.3in]{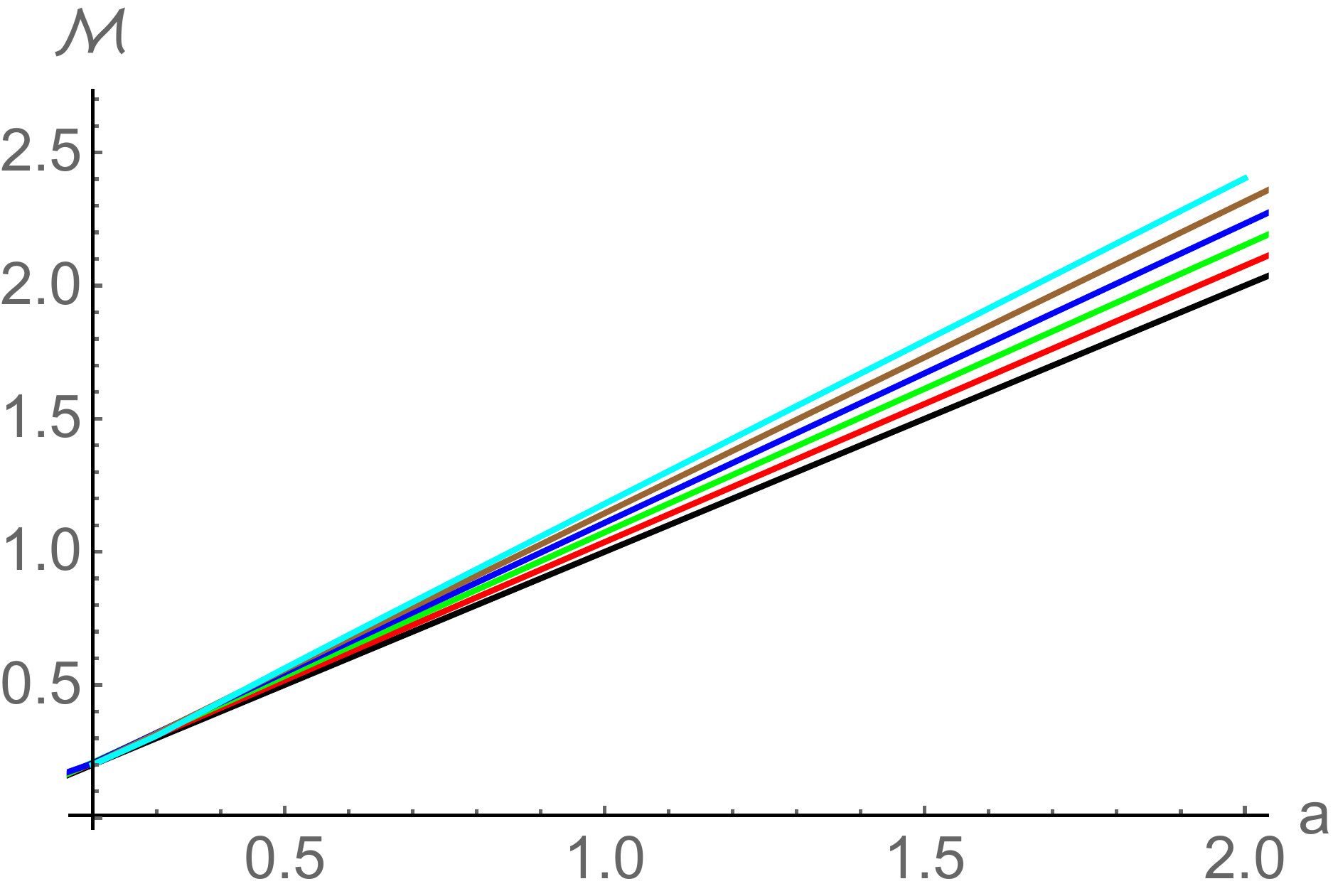}
\caption{\small The behaviour of $\mathcal{M}$ vs $a$ for the extremal case. Here $\ell=0.4$ is used. The black, red, green, blue, brown and cyan curves correspond to $\alpha=0$, $0.2$, $0.4$, $0.6$, $0.8$ and $1.0$ respectively.}
\label{avsMvsalphaLPt4Extreme}
\end{minipage}
\end{figure}

It is interesting to analyse the geometric properties of the rotating hairy solution and contrast them with the Kerr case. The horizon radius can now be obtained by solving for $g^{rr}=0$, which leads to
\begin{eqnarray}
& & a^2 + r_{h}^2 -2 \mathcal{M} r_h + \alpha r_h^2 e^{-r_h/(\mathcal{M}-\alpha \ell/2)}=0 \,.
\label{rotatinghorizon}
\end{eqnarray}
Unfortunately, Eq.~(\ref{rotatinghorizon}) cannot be solved analytically, however, it is straightforward to solve it numerically. We find that, like the Kerr geometry, there are two horizon radii: outer and inner horizons. Both these horizon radii depend nontrivially on $\alpha$ and $\ell$. In particular, the inner horizon radii always increase with $\alpha$, whereas the outer horizon radii decrease with it. This is shown in Figure~\ref{rvsdelvsalphaLPt2aPt1}, where the points where $\Delta(r)$ changes sign indicate the horizon. It is important to point out one subtlety that one can also consider some particular values of $\ell$ for which the outer horizon radii also increase with $\alpha$. However, we found that for all such values the strong energy condition gets violated outside the outer horizon. For the Kerr geometry, we must have $M =\mathcal{M} \geq a$ to avoid the naked singularity. The equality condition $\mathcal{M}=a$ corresponds to the extremal case where the outer and inner horizon radii become equal. We find that the $\mathcal{M}$ vs $a$ relation for the extremal hairy case depends nontrivially on $\alpha$. In particular, the slope of $a$ vs $\mathcal{M}$ curve is always greater than one. This is illustrated in Figure~\ref{avsMvsalphaLPt4Extreme}. This indicates that the maximum angular velocity that the hairy black hole can attain to avoid the naked singularity depends nontrivially on $\alpha$, and in particular, one should always have $\mathcal{M} > a$ in order to mask the singularity for the hairy case.

\begin{figure}[h!]
\begin{minipage}[b]{0.5\linewidth}
\centering
\includegraphics[width=2.8in,height=2.3in]{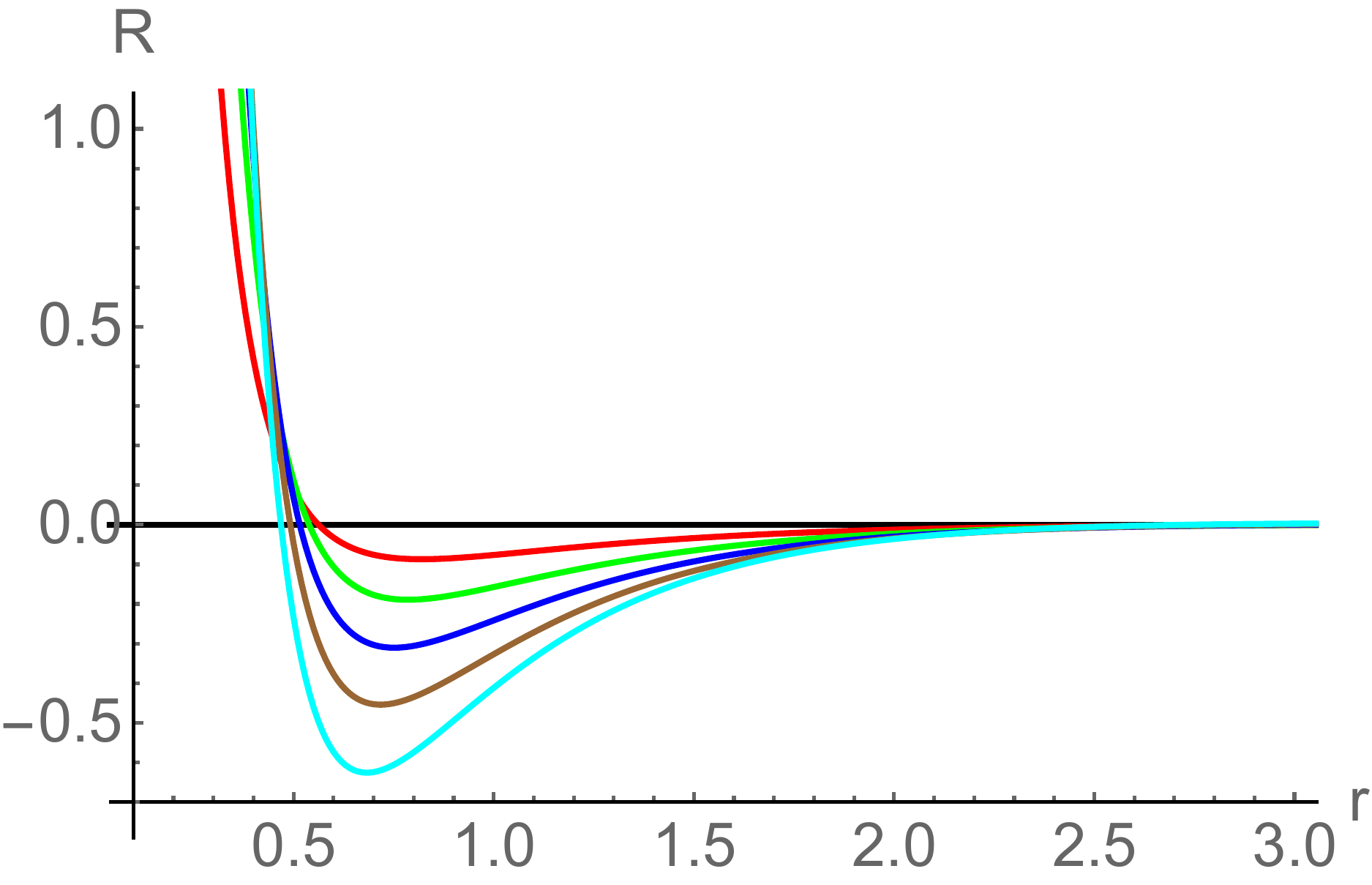}
\caption{ \small The profile of Ricci scalar for different values of $\alpha$ in the plane $\theta=\pi/2$. Here $\mathcal{M}=1$, $\ell=0.4$, and $a=0.1$ are used. The black, red, green, blue, brown and cyan curves correspond to $\alpha=0$, $0.2$, $0.4$, $0.6$, $0.8$, and $1.0$ respectively. }
\label{rvsRiccivsalphaM1aPt1}
\end{minipage}
\hspace{0.4cm}
\begin{minipage}[b]{0.5\linewidth}
\centering
\includegraphics[width=2.8in,height=2.3in]{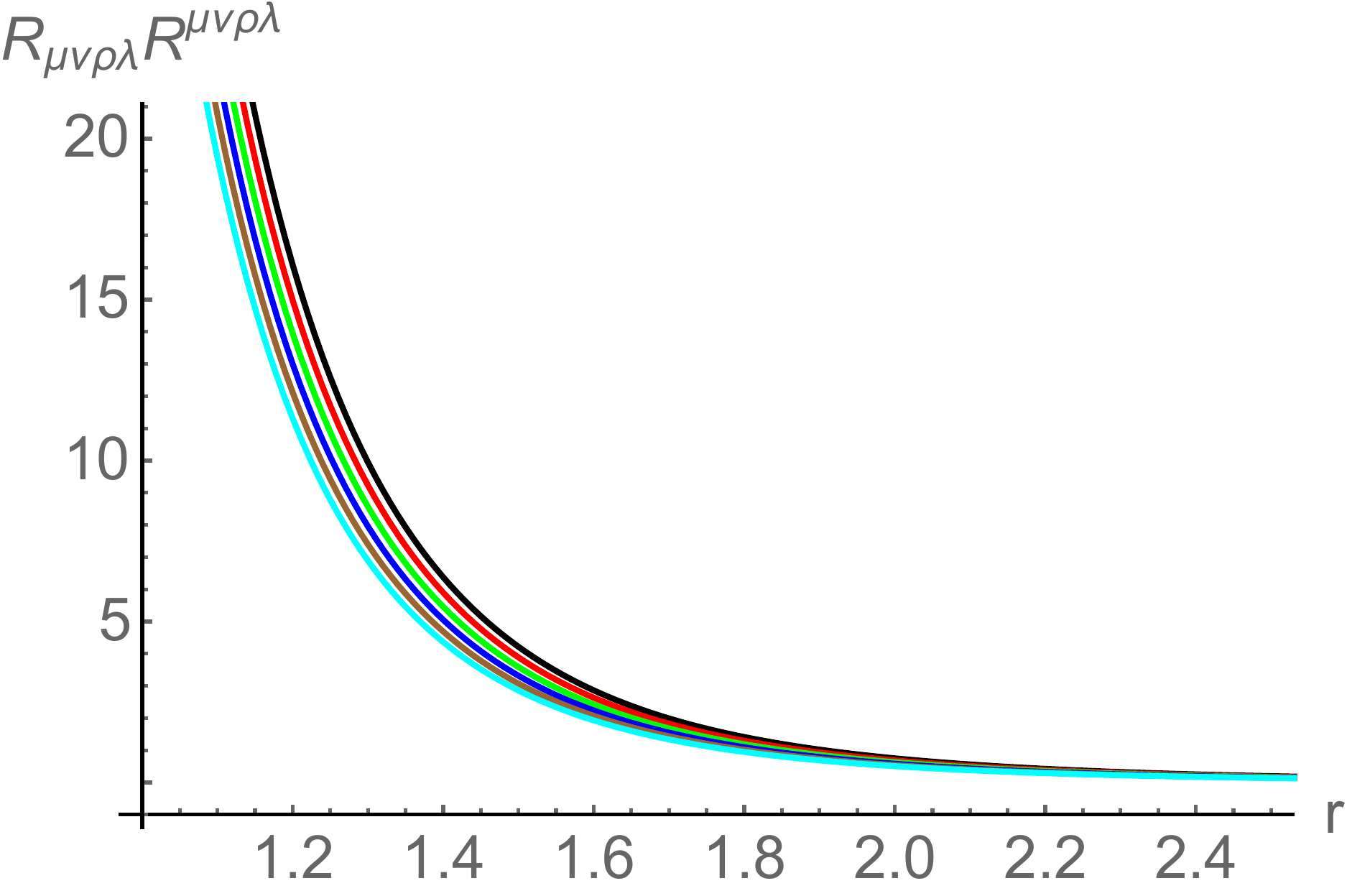}
\caption{\small The profile of Kretschmann scalar for different values of $\alpha$ in the plane $\theta=\pi/2$. Here $\mathcal{M}=1$, $\ell=0.4$, and $a=0.1$ are used. The black, red, green, blue, brown and cyan curves correspond to $\alpha=0$, $0.2$, $0.4$, $0.6$, $0.8$, and $1.0$ and respectively.}
\label{rvsKretschmannvsalphaM1aPt1}
\end{minipage}
\end{figure}

We can further evaluate the scalar invariants. The Ricci scalar is proportional to $\Sigma^{-1}$ and has a simple form
\begin{eqnarray}
& & R = \frac{\alpha \left( 2 M^2 - 4 M r + r^2\right)}{M^2 \Sigma} \,,
\label{Ricci}
\end{eqnarray}
whereas the Kretschmann scalar $R_{\mu\nu\rho\lambda}R^{\mu\nu\rho\lambda}$ is proportional to $\Sigma^{-6}$. The exact expression of the Kretschmann scalar is too long to reproduce here and is not very illuminating, therefore we skip reproducing it here. The rotating hairy solution exhibits a ring singularity in the plane $\theta=\pi/2$ at $r=0$. Importantly, there are no additional singularities in the hairy case than those already present in the Kerr case. Near $r=0$, both invariants have positive curvatures. This is shown in Figures \ref{rvsRiccivsalphaM1aPt1} and \ref{rvsKretschmannvsalphaM1aPt1}. Interestingly, there can be special points in the hairy spacetime where the Ricci scalar vanishes. This happens at $r=2M\pm \sqrt{2}M$, when the numerator of (\ref{Ricci}) goes to zero. These points are moreover $\alpha$ independent, as is shown in Figure~\ref{rvsRiccivsalphaM1aPt1}.

\begin{figure}[h!]
\centering
\includegraphics[width=2.8in,height=2.3in]{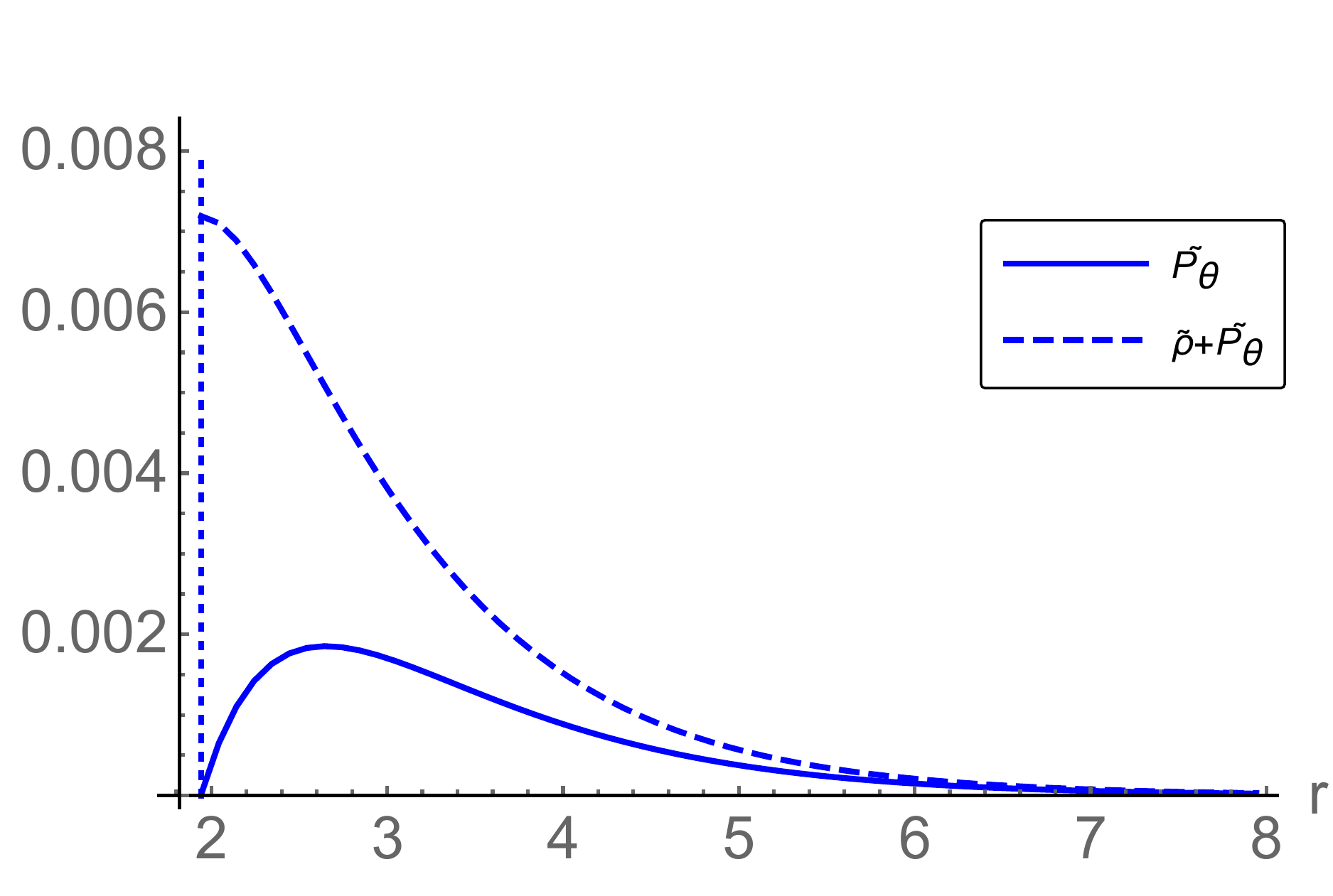}
\caption{\small The profile of $\tilde{p}_\theta$ and $\tilde{\rho}+\tilde{p}_\theta$. Here $\{\mathcal{M}=1,\ell=0.3,\alpha=0.2\}$ are used. The vertical dashed line corresponds to the horizon radii.}
\label{rvsSECrotating}
\end{figure}

From the Einstein tensor, we can further evaluate the density and pressure components of the energy momentum tensor in the rotating case. These are given by,
\begin{eqnarray}
& &  \tilde{\rho} = \frac{r^2 \left(1-e^{-2\lambda}-r (e^{-2\lambda})' \right)}{\kappa^2\Sigma^2}= \frac{\alpha r^2 e^{-\frac{r}{M}}}{\kappa^2 M \Sigma^2}(r-M)\,, \nonumber \\
& & \tilde{p}_r =- \tilde{\rho} \,, \nonumber \\
& & \tilde{p}_\theta = -\tilde{p}_r + \frac{1}{\kappa^2\Sigma} \left(\frac{r^2 (e^{-2\lambda})''}{2}+2 r (e^{-2\lambda})'+(e^{-2\lambda})-1 \right) \nonumber \\
& &  ~~~= \frac{\alpha r^2 e^{-\frac{r}{M}}}{\kappa^2 M \Sigma^2}(r-M) + \frac{\alpha e^{-\frac{r}{M}}}{2\kappa^2 M^2 \Sigma}(2M^2-4Mr+r^2) \,,  \nonumber \\
& & \tilde{p}_\varphi = \tilde{p}_\theta \,.
\label{SETrotational}
\end{eqnarray}
We can immediately notice that these expressions reduce to the standard static expressions (\ref{densitypressureSEC}) when $a\rightarrow 0$ or $\theta=\pi/2$. The physical ring singularity nature of the spacetime at $r=0$ is further reflected in the density and pressure components. Notice that these components, written in terms of ADM mass $\mathcal{M}$, depend on the parameter $\ell$. This allows us to choose suitable values of hairy parameters $\alpha$ and $\ell$ such that the strong energy condition is satisfied everywhere outside the horizon in the rotating case as well. For example, for $\{\mathcal{M}=1,\alpha=0.2,a=0.1\}$, we must have $\ell\gtrsim 0.3$ to satisfy the strong energy condition. This is shown in Figure~\ref{rvsSECrotating}, where we can explicitly see that both $\tilde{p}_\theta$ and $\tilde{\rho}+\tilde{p}_\theta$ are positive outside the horizon. In general, compared to the static spacetime, a higher value of $\ell$ is needed to satisfy the strong energy condition when the rotation parameter $a$ increases.

Here we like to end this section by mentioning that the axisymmetric solution can also be obtained using the standard Newman-Janis algorithm. However, this algorithm involves a certain level of arbitrariness in the  complexification of null tetrads. Indeed, most of the time, there is no unique way of complexifying all the terms in the tetrad in the same way. We can still, however, generate a consistent rotating hairy metric in our case using the Newman-Janis algorithm. This required choosing a particular form of the complexified null tetrads. In Appendix B, we will briefly mention these complexified null tetrads for the hairy black hole.

\section{Thermodynamics of the rotating hairy solution}
\label{S4}
Having obtained the rotating solution of the hairy black hole, we now move on to discuss its thermodynamic properties. For this purpose, we first recast the metric elements of (\ref{rotatinghairmet}) in the following form
\begin{eqnarray}
& & g_{tt} = \frac{\Delta-a^2\sin^2\theta}{\Sigma}\,,~~~g_{rr}=-\frac{\Sigma}{\Delta}\,,~~~g_{\theta\theta}=-\Sigma \,,\nonumber \\
& & g_{t\varphi} = \frac{a \sin^2\theta (r^2 + a^2 -\Delta)}{\Sigma}\,,~~~ g_{\varphi\varphi} = -\frac{\sin^2\theta \left((r^2 + a^2)^2 -\Delta a^2 \sin^2\theta \right)}{\Sigma} \,, \nonumber \\
& & \Delta(r)= a^2 + r^2 -2 \mathcal{M} r + \alpha r^2 e^{-r/(\mathcal{M}-\alpha \ell/2)} \,.
\label{rotatinghairmet1}
\end{eqnarray}
The Bekenstein-Hawking entropy of the black hole can be computed from the area of the horizon, which for the above metric is given by
\begin{eqnarray}
& & S_{BH} = \frac{1}{4 G_N}\int d\theta d\varphi \sqrt{g_{\theta\theta}g_{\varphi\varphi}} = \int d\theta d\varphi \sin\theta \sqrt{\left((r^2 + a^2)^2 -\Delta a^2 \sin^2\theta \right)} \,,
\end{eqnarray}
since at the horizon $\Delta(r_h)=0$, we get
\begin{eqnarray}
& & S_{BH} = \frac{\pi \left(r_{h}^{2}+a^2\right)}{G_N} \,.
\label{rotatinghairyentropy}
\end{eqnarray}
Notice that in $S_{BH}$ the hairy parameter $\alpha$ does not appear explicitly, implying that the hairy black hole entropy expression remains the same as in the non-hairy Kerr case. However, $S_{BH}$ does depend implicitly on $\alpha$ through $r_h$.

Similarly, the temperature of the rotating black hole can be obtained from the standard Euclidean procedure \cite{Gibbons:1976ue}. For this purpose, we follow the procedure outlined in \cite{Mann:1996bi} and note that the above spacetime enjoys a pair of orthogonal vectors
\begin{eqnarray}
& & \xi = \partial_t + \frac{a}{a^2+r^2}\partial_\varphi\,,~~~ \tilde{\xi}= a \sin^2\theta \partial_t + \partial_\varphi \,.
\label{lkillingvec}
\end{eqnarray}
The vector $\xi$ is null on the horizon $\xi^2(r_h)=0$ and is time-like everywhere outside the horizon $r\geq r_h$, whereas the vector $\tilde{\xi}$ is spacelike for $r\geq r_h$.  The corresponding one-forms dual to $\xi$ and $\tilde{\xi}$ are
\begin{eqnarray}
& & d\omega = \frac{a^2+r^2}{\Sigma}(dt-a\sin^2\theta d\varphi),~~~ d\tilde{\omega}= \frac{a^2+r^2}{\Sigma}(-\frac{a}{a^2+r^2}dt+d\varphi) \,.
\end{eqnarray}
We now consider the Euclideanization of the rotating hairy metric. The standard procedure \cite{Gibbons:1976ue} suggests the change of time variable $t=i \tau$, along with the transformation of the rotating parameter $a=i\hat{a}$. With these transformations, the Euclidean vectors $\xi$ and $\tilde{\xi}$ and their one forms $d\omega$ and $d\tilde{\omega}$ reduce to
\begin{eqnarray}
& & \xi = \partial_\tau - \frac{\hat{a}}{r^2-\hat{a}^2}\partial_\varphi\,,~~~ \tilde{\xi}= \hat{a} \sin^2\theta \partial_\tau + \partial_\varphi \,, \nonumber \\
& & d\omega = \frac{r^2-\hat{a}^2}{\hat{\Sigma}}(d\tau-\hat{a}^2\sin^2\theta d\varphi),~~~ d\tilde{\omega}= \frac{r^2 - \hat{a}^2}{\hat{\Sigma}}\left(\frac{\hat{a}}{r^2-\hat{a}^2}d\tau+d\varphi\right) \,,
\end{eqnarray}
where $\hat{\Sigma}=r^2-\hat{a}^2\cos^2\theta$. The Euclidean metric now simply reduces to
\begin{eqnarray}
& & ds_{E}^2 = - \frac{\hat{\Delta}\hat{\Sigma}}{\left(r^2-\hat{a}^2 \right)^2}d\omega^2 - \frac{\hat{\Sigma}}{\hat{\Delta}} dr^2 - \hat{\Sigma} \left(d\theta^2 + \sin^2\theta d\tilde{\omega}^2   \right) \,,
\label{Euclideanrotmet}
\end{eqnarray}
where $\hat{\Delta}=r^2-\hat{a}^2-2\mathcal{M}r+\alpha r^2 e^{-r/(\mathcal{M}-\alpha \ell/2)}$, which also give the radius of the horizon $r=\hat{r}_h$. Introducing a new radial coordinate $x$ near the horizon
\begin{eqnarray}
& & \hat{\Delta}= \eta (r-\hat{r}_h) = \frac{\eta^2 x^2}{4}\,, \nonumber \\
& & \eta = \frac{\hat{r}_{h}^2 + \hat{a}^2}{\hat{r}_{h}} + \alpha \hat{r}_{h} e^{-\frac{\hat{r}_{h}}{\mathcal{M}-\alpha \ell/2}} \left(1- \frac{\hat{r}_{h}}{\mathcal{M}-\alpha \ell/2} \right)\,,
\end{eqnarray}
the Euclidean metric (\ref{Euclideanrotmet}) near the horizon takes the form
\begin{eqnarray}
& & ds_{E}^2 = - \hat{\Sigma}_h\left(dx^2 + \frac{\eta^2 x^2}{4(\hat{r}_{h}^2 - \hat{a}^2)^2}d\omega^2\right) - \hat{\Sigma}_h \left(d\theta^2 + \sin^2\theta d\tilde{\omega}^2   \right) \,,
\label{Euclideanrotmet1}
\end{eqnarray}
where $\hat{\Sigma}_h=\hat{r}_{h}^2-\hat{a}^2\cos^2\theta$. On the horizon surface $\mathcal{H}$, one can further introduce a well defined coordinate $\psi=\varphi+\frac{\hat{a} \tau}{\hat{r}_{h}^2-\hat{a}^2}$, giving metric on the horizon
\begin{eqnarray}
& & ds_{\mathcal{H}}^2 = \hat{\Sigma}_h \left(d\theta^2 + \sin^2\theta d\tilde{\omega}^2 \right) = \hat{\Sigma}_{h} \left(d\theta^2 + \frac{\left(\hat{r}_{h}^2-\hat{a}^2\right)^2}{\hat{\Sigma}_{h}^2} \sin^2\theta d\psi^2 \right)  \,.
\label{Euclideanhorizonmet}
\end{eqnarray}
Notice that the points $\psi$ and $\psi+2\pi$ should be identified on the horizon $\mathcal{H}$ for the regularity of the metric (\ref{Euclideanhorizonmet}) at points $\theta=0$ and $\theta=\pi$. The metric (\ref{Euclideanrotmet1}) can be rewritten as
\begin{eqnarray}
& & ds_{E}^2 = - \hat{\Sigma}_h ds_{C_2}^2 - ds_{\mathcal{H}}^2  \,,
\label{Euclideanrotmet2}
\end{eqnarray}
where $ds_{C_2}^2$ is the metric on two dimensional disk $C_2$
\begin{eqnarray}
& & ds_{C_2}^2 = dx^2 + \frac{\eta^2 x^2}{4(\hat{r}_{h}^2 - \hat{a}^2)^2}d\omega^2  \,.
\label{Euclideandisk}
\end{eqnarray}
Considering the above metric with fixed $(\theta,\psi)$ and introducing a new angle coordinate $\chi=\tau-\hat{a}\sin^2\theta \varphi$, the metric on disk takes the form
\begin{eqnarray}
& & ds_{C_2}^2 = dx^2 + \frac{\eta^2 x^2}{4\hat{\Sigma}_{h}^2}d\chi^2 = dx^2 + x^2 d \left(\frac{\eta}{2\hat{\Sigma}_{h}}\chi \right)^2  \,.
\label{Euclideandisk2}
\end{eqnarray}
This metric takes the same form as a plane in polar coordinate if $\eta/(2\hat{\Sigma}_{h})$ has the period $2\pi$. Otherwise, there is a conical singularity at $x=0$. One must therefore identify the points $\chi$ and $\chi+4\pi\hat{\Sigma}_{h}\eta^{-1}$ to avoid the conical singularity. Since this must be true independently of the coordinate $\theta$ on the horizon $\mathcal{H}$, one must also identify the points $(\tau, \varphi)$ with $(\tau+\beta_h, \varphi - \Omega_h \beta_h)$, where $\beta_h=4\pi\hat{\Sigma}_{h}/\eta$ is inverse Hawking temperature of the black hole and $\Omega=\hat{a}/(\hat{r}_{h}^2-\hat{a}^2)$ is the angular velocity. Now, analytically continuing the results back to the real values $a$, we have the temperature and angular velocity of the hairy rotating black hole as
\begin{eqnarray}
& & T_h= \frac{1}{\beta_h} = \frac{r_{h}^2-a^2+\alpha r_{h}^2 e^{-\frac{r_h}{\mathcal{M}-\alpha \ell/2}}\left(1-\frac{r_h}{\mathcal{M}-\alpha \ell/2}\right)}{4 \pi r_h \left(r_{h}^2 + a^2 \right)}  \,, \nonumber \\
& & \Omega_h = \frac{a}{r_{h}^2+a^2} \,.
\label{Temprothairy}
\end{eqnarray}
Note that the Hawking temperature gets a correction due to hair, and it reduces to the standard Kerr temperature when the hair parameter $\alpha$ goes to zero. The expression of angular velocity, like the entropy, does not depend explicitly on $\alpha$ and remains the same as in the Kerr case. Therefore, the angular velocity of the horizon is again given by $\Omega_h=-g_{t\varphi}/g_{\varphi\varphi}$, as is expected. The angular momentum of the black hole is similarly given by
\begin{eqnarray}
& & J=a \mathcal{M} \,.
\label{Angularmomentum}
\end{eqnarray}

We now discuss the notion of quasilocal energy in the hairy rotating black hole and its relation with the mass parameter $\mathcal{M}$. For this purpose, we use the Brown-York definition of quasilocal energy \cite{Brown:1992br}.  To obtain it, one first considers foliations of four-dimensional manifold $\mathcal{N}$ by spacelike constant time hypersurfaces $\Gamma$. The spacetime is spatially bounded by the three-dimensional surface $B^{(3)}$ ($r=$ constant surface), and the intersection of $\Gamma$ with the boundary $B^{(3)}$ is a two-dimensional surface $B^{(2)}$ with induced metric $\sigma_{ab}$, i.e. $B^{(2)}$ is a $r=$ constant hypersurface
within $\Gamma$. The quasilocal energy is then given as the proper surface integral
\begin{eqnarray}
& & E=\varepsilon - \varepsilon_0 = \frac{1}{\kappa^2} \int _{B^{(2)}} \sqrt{\sigma} (\mathcal{K}-\mathcal{K}^0) \,,
\label{QLEdef}
\end{eqnarray}
where $\mathcal{K}$ is the trace of the extrinsic curvature of $B^{(2)}$ as embedded in $\Gamma$ and $\mathcal{K}^0$ represents the trace of extrinsic curvature for some reference spacetime \footnote{The $\mathcal{K}$ here should not be confused with the $K$ appeared in the previous section.}. To compute the quasilocal energy for asymptotically flat solutions, the reference space is generally taken to be a flat three-dimensional slice $E^{(3)}$ of flat spacetime. For our hairy black hole, the trace of the extrinsic curvature $\mathcal{K}$ of $r=$ constant surface within the $t=$ constant hypersurface is given by
\begin{eqnarray}
& & \mathcal{K} = -\sqrt{\frac{\Delta}{\Sigma}}\frac{\partial_r \left[(r^2+a^2)^2-\Delta a^2 \sin^2\theta \right]}{2\left[(r^2+a^2)^2-\Delta a^2 \sin^2\theta \right]} \,, \nonumber\\
& & = - \frac{\sqrt{r^2-2\mathcal{M} r+a^2+\alpha r^2 e^{-\frac{r}{M}}}}{2\sqrt{r^2+a^2\cos^2\theta}} \frac{\left[4r(r^2+a^2)-a^2\sin^2\theta \Delta'(r) \right] }{\left[(r^2+a^2)^2-\Delta a^2 \sin^2\theta \right]}\,.
\label{Kunref}
\end{eqnarray}
Using $\sqrt{\sigma}=\sin\theta \sqrt{(r^2+a^2)^2-\Delta a^2 \sin^2\theta}$ for the induced metric on the two-sphere $B^{(2)}$, the unreferenced Brown-York energy
for the hairy rotating black hole within a sphere of radius $r$ can be evaluated as
\begin{eqnarray}
& & \varepsilon = \frac{1}{8 \pi} \int  \sqrt{\sigma} \mathcal{K} = \frac{1}{8 \pi} \int d\theta d\varphi  \sin\theta \sqrt{(r^2+a^2)^2-\Delta a^2 \sin^2\theta} \mathcal{K} \,, \nonumber \\
& & = - \frac{r}{8} \sqrt{1-\frac{2 \mathcal{M}}{r}+ \frac{a^2}{r^2}+\alpha e^{-\frac{r}{M}}} \int_{0}^{\pi} d\theta \sin\theta\frac{\left[4r(r^2+a^2)-a^2\sin^2\theta \Delta'(r) \right]}{\sqrt{\left(r^2+a^2\cos^2\theta \right) \left[(r^2+a^2)^2-\Delta a^2 \sin^2\theta \right]}} \,. \nonumber \\
\label{Kunref1}
\end{eqnarray}
Unfortunately, the above equation in general can not be integrated in closed form. Therefore, one has to rely on some approximation for which the above integration can be performed. A physically motivated and particularly interesting approximation is a slow rotation limit i.e. $a/r\ll 1$, in which case a closed form expression can be obtained. In this approximation, the unreferenced Brown-York energy reduces to
\begin{eqnarray}
& &  \varepsilon = - \frac{r}{2} \sqrt{1-\frac{2 \mathcal{M}}{r}+ \frac{a^2}{r^2}+\alpha e^{-\frac{r}{M}}} \int_{0}^{\pi} d\theta \sin\theta \left[1 - \frac{a^2 \cos^2\theta}{2r^2} + \frac{a^2 \sin^2\theta}{2r^2} \left( -\frac{\mathcal{M}}{r} + \frac{\alpha r e^{-\frac{r}{M}}}{2M}  \right) \right] \,, \nonumber\\
& &  =  - r \sqrt{1-\frac{2 \mathcal{M}}{r}+ \frac{a^2}{r^2}+\alpha e^{-\frac{r}{M}}} \left[1 - \frac{a^2}{6 r^2} \left(1 + \frac{2\mathcal{M}}{r} - \frac{\alpha r e^{-\frac{r}{M}}}{M}    \right)  \right]\,.
\label{Kunrefslowrot}
\end{eqnarray}
Notice that for large $r$, we have $\varepsilon\rightarrow \mathcal{M}-r$, which is divergent when $r \rightarrow \infty$. This suggests the need for a subtraction term, i.e the quasilocal energy of a reference spacetime, to renormalize the unreferenced quasilocal energy. Obtaining the Brown-York energy for reference spacetime is, however, more difficult \cite{Martinez:1994ja,Bose:1999er,Chakraborty:2015kva}. For this one first needs to find a two-dimensional surface isometric to $B^{(2)}$, which is consistently embedded in a three-dimensional flat space $E^{(3)}$. If this $E^{(3)}$ is described by the coordinates $\{\mathcal{R}, \Theta, \Phi \}$, then
\begin{eqnarray}
& & ds^2= d\mathcal{R}^2 +\mathcal{R}^2 d\Theta^2 + \mathcal{R}^2 \sin^2\Theta d\Phi^2\,.
\label{refmet}
\end{eqnarray}
The desired two-dimensional surface isometric to $B^{(2)}$ can be defined by $\mathcal{R}=f(\Theta)$, where $f$ is a function of the azimuthal angle $\Theta$. Its intrinsic metric or curvature can be evaluated from (\ref{refmet}). For this purpose, we first assume that $\Theta=\Theta(\theta)$ and $\Phi=\varphi$, which makes $\mathcal{R}$ is a function of $\theta$ on the two-surface. The line element on it takes the form
\begin{eqnarray}
& & ds^2=\left[\dot{\mathcal{R}}^2 +\mathcal{R}^2 \dot{\Theta}^2 \right] + \mathcal{R}^2 \sin^2\Theta d\Phi^2\,,
\label{refmet1}
\end{eqnarray}
where an overdot denotes derivative with respect to $\theta$. Requiring the above line element to be isometric to (\ref{rotatinghairmet1}) gives us the following coupled differential equations for $\mathcal{R}$ and $\Theta$,
\begin{eqnarray}
& & \dot{\mathcal{R}}^2 + \mathcal{R}^2 \dot{\Theta}^2 = r^2+a^2\cos^2\theta  \,, \nonumber\\
& & \mathcal{R}^2 \sin^2\Theta  = \frac{\sin^2\theta \left((r^2 + a^2)^2 -\Delta a^2 \sin^2\theta \right)}{r^2+a^2\cos^2\theta} \,.
\label{KrefREOM}
\end{eqnarray}
These two coupled equations need to be solved to get $\mathcal{R}$ and $\Theta$ in terms of $\theta$. Unfortunately, these equations are again difficult to solve in a closed form for a general value of $a$. Therefore, once again we resort to the slow rotating approximation to get the analytic results. In this approximation,  the above equations can be combined to obtain the following first-order equation \footnote{Here the $\dot{\mathcal{R}}^2 \simeq O(a^4/r^4)$ term is neglected in the calculation.}
\begin{eqnarray}
& & \frac{d\Theta}{\sin\theta}=\frac{d\theta}{\sin\theta}\left[1-\frac{a^2\sin^2\theta}{2 r^2}\left(1+\frac{2\mathcal{M}}{r} - \alpha e^{-\frac{r}{M}} \right) \right]\,,
\end{eqnarray}
which can be solved to give
\begin{eqnarray}
& & \sin\Theta = \sin\theta \left[1+ \frac{a^2 \cos^2\theta}{2 r^2} \left(1 + \frac{2 \mathcal{M}}{r} -\alpha e^{-\frac{r}{M}}  \right)  \right]  \,.
\label{KrefRsoltheta}
\end{eqnarray}
Putting this back into (\ref{KrefREOM}) and simplifying, we can obtain the solution for $\mathcal{R}$
\begin{eqnarray}
& & \mathcal{R}= r \left[1+ \frac{a^2 \sin^2\theta}{2 r^2} -\frac{a^2 \cos2\theta}{2 r^2} \left(\frac{2\mathcal{M}}{r} - \alpha e^{-\frac{r}{M}}   \right)     \right] \,.
\label{KrefRsolR}
\end{eqnarray}
From these equations, we can also find $\mathcal{R}$ in terms of $\Theta$
\begin{eqnarray}
& & \mathcal{R}= r \left[1+ \frac{a^2 \sin^2\Theta}{2 r^2} -\frac{a^2 \cos2\Theta}{2 r^2} \left(\frac{2\mathcal{M}}{r} - \alpha e^{-\frac{r}{M}}   \right)     \right] \,,
\label{KrefRsolR}
\end{eqnarray}
which would allow us to write the intrinsic metric on two-surface, as embedded in flat space, as
\begin{eqnarray}
 ds^2 \approx r^2 \left[1 + \frac{a^2 \sin^2\Theta}{r^2} - \frac{a^2 \cos2\Theta}{ r^2} \left(\frac{2\mathcal{M}}{r} - \alpha e^{-\frac{r}{M}}   \right)    \right] \left(d\Theta^2 + \sin^2\Theta d\Phi^2  \right) \,.
\label{indmettwosurface}
\end{eqnarray}
This metric describes a surface in flat three-dimensional space $E^{(3)}$ whose intrinsic curvature equals the intrinsic curvature of a $r=$ constant surface in the rotating hairy space. It can be used to compute the extrinsic curvature of the reference flat spacetime, which in turn can be used to compute the reference contribution to the quasilocal energy in Eq.~(\ref{QLEdef}). A long but direct computation in the slow rotating approximation gives us \footnote{The general expressions to compute the trace of the extrinsic curvature $\mathcal{K}^0$, corresponding to two-dimensional surfaces embedded in a three-dimensional axisymmetric space, can be found in \cite{Martinez:1994ja,Bose:1999er}.}
\begin{eqnarray}
& & \varepsilon_0 = \frac{1}{8\pi}\int d\Theta d\Phi \sqrt{\sigma} \mathcal{K}^0 = -r \left[1+ \frac{a^2}{3 r^2} \left(1+ \frac{\mathcal{M}}{r} - \frac{\alpha e^{-\frac{r}{M}}}{2}   \right)   \right]  \,.
\label{Kref}
\end{eqnarray}
The quasilocal energy obtained by subtraction (\ref{Kref}) from (\ref{Kunrefslowrot}). The result is
\begin{eqnarray}
 E= r\left[1-\sqrt{1-\frac{2\mathcal{M}}{r}+\frac{a^2}{r^2}+\alpha e^{-\frac{r}{M}}} \right] +  \frac{a^2}{3 r} \left(1+ \frac{\mathcal{M}}{r} - \frac{\alpha e^{-\frac{r}{M}}}{2}   \right)  \nonumber \\
   + \frac{a^2}{6 r} \left(1+\frac{2\mathcal{M}}{r} - \frac{ \alpha r e^{-\frac{r}{M}}}{M} \right)\sqrt{1-\frac{2\mathcal{M}}{r}+\frac{a^2}{r^2}+\alpha e^{-\frac{r}{M}}}\,.
\label{QLEsub}
\end{eqnarray}
This gives the quasilocal energy of a slowly rotating hairy spacetime within a sphere of radius $r$. In the asymptotic limit $r\rightarrow \infty$, we have
\begin{eqnarray}
 E \rightarrow \mathcal{M} + \mathcal{O}(a^2/r^2, \alpha r e^{-r/M})\,,
\label{QLEsub}
\end{eqnarray}
indicating that $\mathcal{M}$ is indeed the ADM energy of the hairy system.

We can further use (\ref{QLEsub}) to compute the energy at the horizon (whenever $a\ll r_h$ or $a\ll \mathcal{M}$). In the slow rotating approximation, at $r=r_h$, we have
\begin{eqnarray}
& & E(r=r_h)=r_h \left[1+ \frac{a^2}{3 r_{h}^{2}} \left(1+ \frac{\mathcal{M}}{r_h} - \frac{\alpha e^{-\frac{r_h}{M}}}{2} \right)   \right]\,, \nonumber\\
& & ~~~~~~~~~~~~~~= r_h \left[1+\frac{a^2}{2 r_h^{2}} + \mathcal{O}(a^4/r_{h}^4)  \right]\,.
\label{QLEhorizon}
\end{eqnarray}
Notice that the above expression of horizon energy is exactly similar to the Kerr case \cite{Martinez:1994ja}. Moreover, we have
\begin{eqnarray}
& & E(r=r_h)\simeq \sqrt{\frac{\text{Horizon Area}}{4 \pi}}=2 \mathcal{M}_{irr}\left[1+\mathcal{O}(a^4/r_{h}^4)\right]\,,
\end{eqnarray}
where $\mathcal{M}_{irr}$ is the irreducible mass of the hairy rotating black hole and is proportional to the square root horizon area \cite{Christodoulou:1971pcn}. Therefore, in the hairy rotating case as well, to leading order in the slow rotating approximation, quasilocal energy at the black hole horizon is twice the irreducible mass.

\begin{figure}[h!]
\begin{minipage}[b]{0.5\linewidth}
\centering
\includegraphics[width=2.8in,height=2.0in]{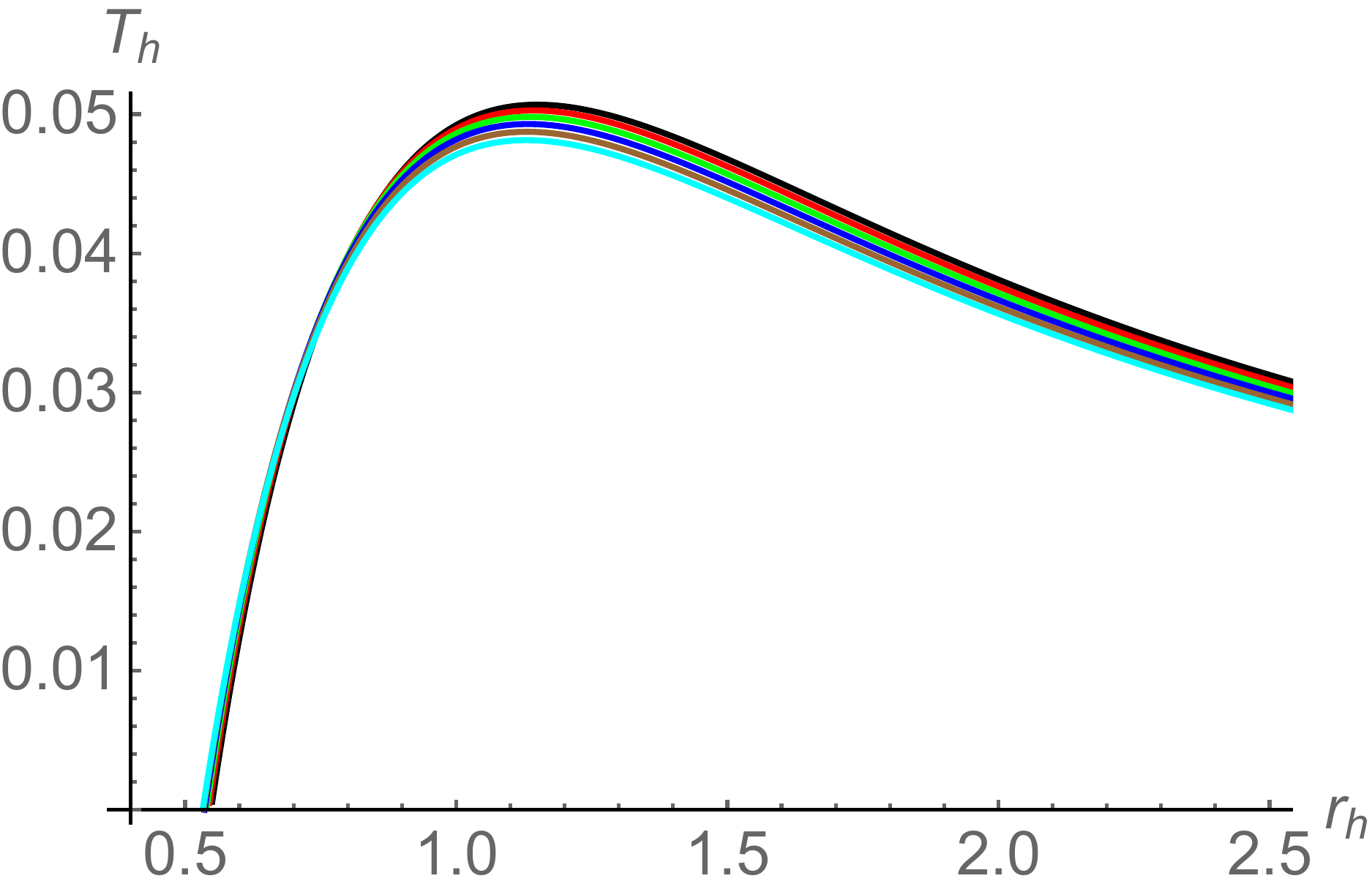}
\caption{ \small The variation of Hawking temperature with respect to the horizon radius for different values of $\alpha$. Here $J=0.3$ and $\ell=0.4$ are used. The black, red, green, blue, brown, and cyan curves correspond to $\alpha=0$, $0.1$, $0.2$, $0.3$, $0.4$, and $0.5$ respectively.}
\label{rhvsTempvsAlphaJPt3LPt4}
\end{minipage}
\hspace{0.4cm}
\begin{minipage}[b]{0.5\linewidth}
\centering
\includegraphics[width=2.8in,height=2.3in]{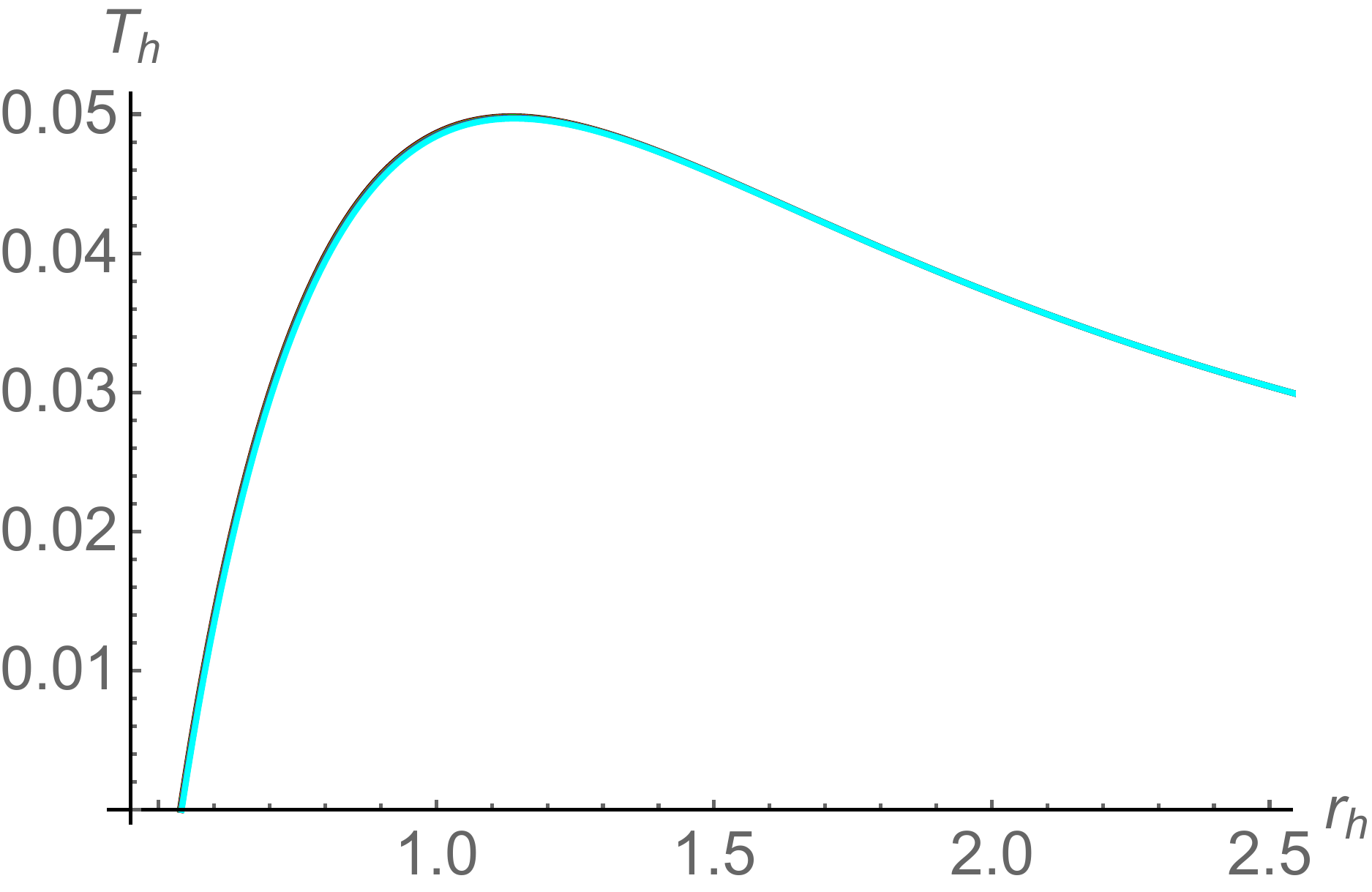}
\caption{\small The variation of Hawking temperature with respect to the horizon radius for different values of $\ell$. Here $J=0.3$ and $\alpha=0.2$ are used. The black, red, green, blue, brown, and cyan curves correspond to $\ell=0.3$, $0.4$, $0.5$, $0.6$, $0.7$ and $0.8$ respectively.}
\label{rhvsTempvsLJPt3AlphaPt2}
\end{minipage}
\end{figure}

We now turn our attention to analysing the thermodynamic properties and stability of the rotating hairy black hole. Here we mainly concentrate on the fixed angular momentum canonical ensemble. The variation of Hawking temperature with respect to the horizon radius for different values of $\alpha$ and $\ell$ is shown in Figures \ref{rhvsTempvsAlphaJPt3LPt4} and \ref{rhvsTempvsLJPt3AlphaPt2}. Notice that for finite $\alpha$ and $\ell$ there are two black hole branches at a fixed temperature. The first branch for which the temperature increases with radius is stable and appears for small $r_h$, whereas the second branch for which the temperature decreases with radius is unstable and appears for large $r_h$. The thermodynamic stable/unstable nature of the small/large black hole branches will be further confirmed from the free energy and specific heat analysis shortly.

The thermodynamic variation of the hairy rotating black hole remains qualitatively similar to that of the Kerr case. In particular, with finite $\alpha$, there are again two black hole branches, with the smaller branch being the stable one. Also, there exists a maximum temperature above which the black hole solution does not exist. This maximum temperature is moreover a hair-dependent quantity. The complete dependence of this maximum temperature on the hair parameters will be discussed shortly. Notice also that the $r_h$ vs $T_h$ profiles for different values of $\ell$ almost overlap with each other, suggesting a limited thermodynamic dependence of hairy black holes on $\ell$.

\begin{figure}[h!]
\begin{minipage}[b]{0.5\linewidth}
\centering
\includegraphics[width=2.8in,height=2.0in]{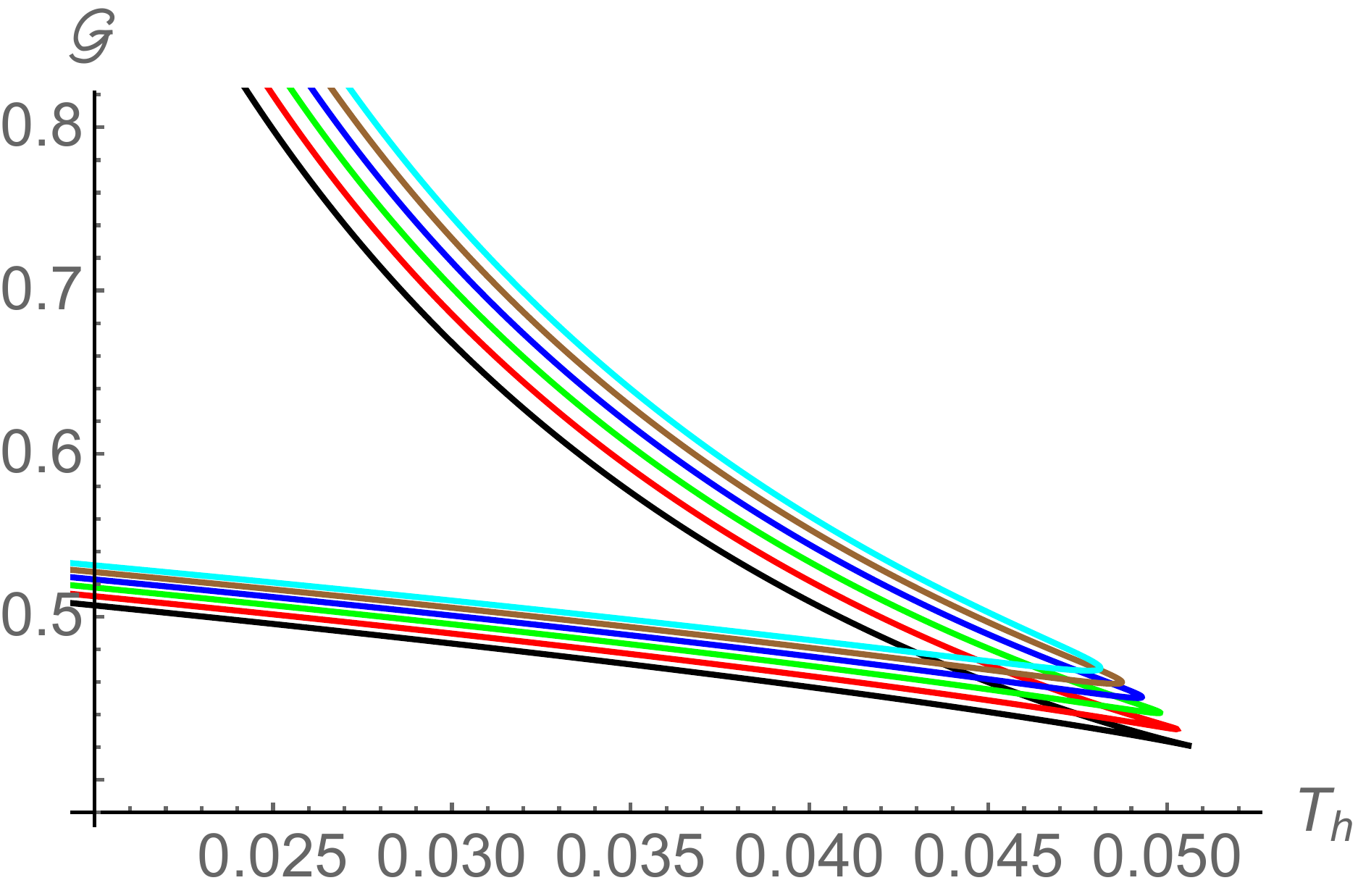}
\caption{ \small The variation of free energy $\mathcal{G}$ with respect to the Hawking temperature for different values of $\alpha$. Here $J=0.3$ and $\ell=0.4$ are used. The black, red, green, blue, brown, and cyan curves correspond to $\alpha=0$, $0.1$, $0.2$, $0.3$, $0.4$, and $0.5$ respectively. In all these curves, the horizon radius $r_h$ increases from left to right and then up.}
\label{TempvsGibbsvsAlphaJPt3LPt4}
\end{minipage}
\hspace{0.4cm}
\begin{minipage}[b]{0.5\linewidth}
\centering
\includegraphics[width=2.8in,height=2.3in]{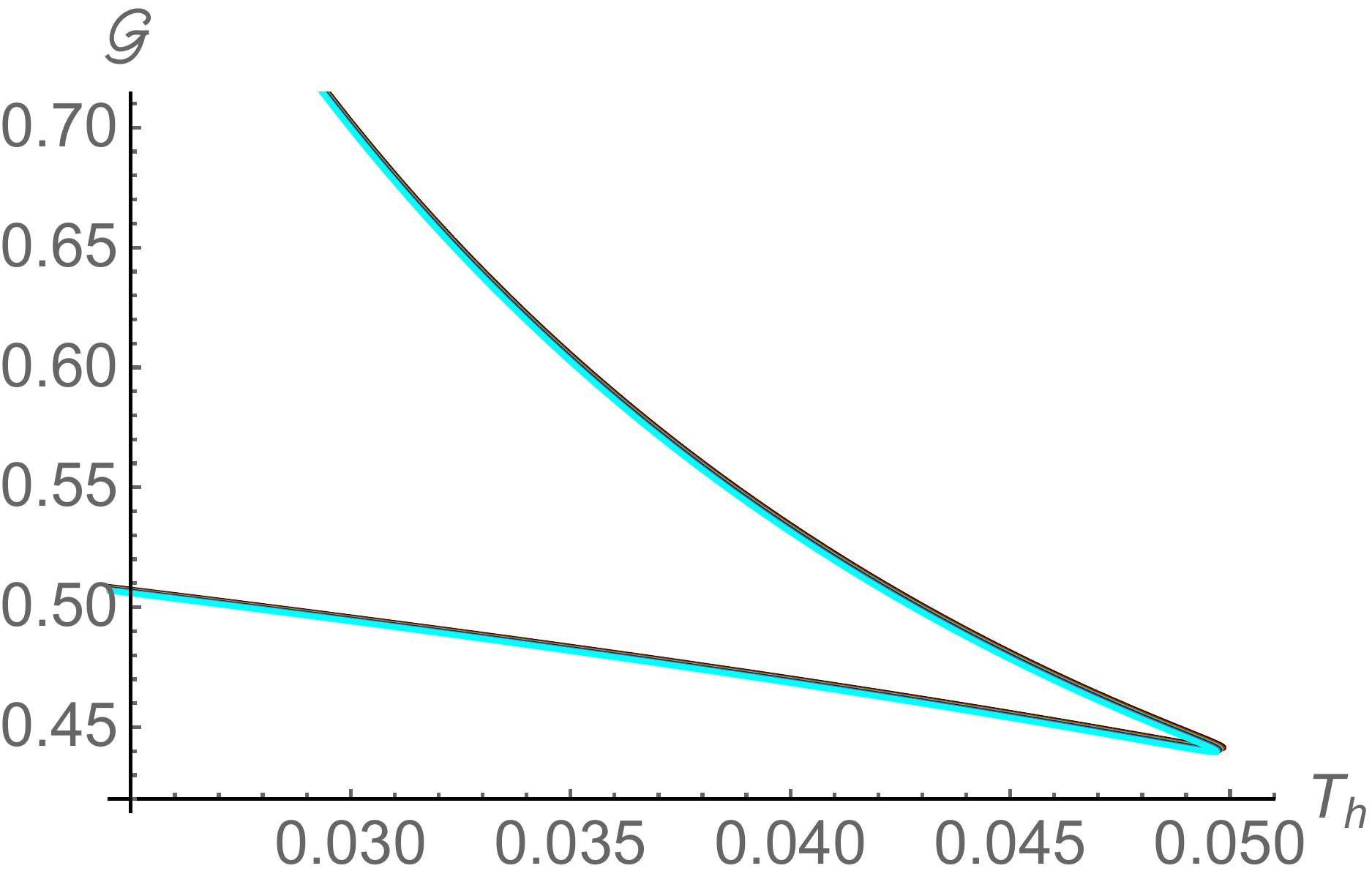}
\caption{\small The variation of free energy $\mathcal{G}$ with respect to the Hawking temperature for different values of $\ell$.  Here $J=0.3$ and $\alpha=0.2$ are used. The black, red, green, blue, brown, and cyan curves correspond to $\ell=0.3$, $0.4$, $0.5$, $0.6$, $0.7$, and $0.8$ respectively. In all these curves, the horizon radius $r_h$ increases from left to right and then up.}
\label{TempvsGibbsvsLJPt3AlphaPt2}
\end{minipage}
\end{figure}

The free energy, again in the fixed angular momentum canonical ensemble, can further disclose the equilibrium thermodynamic stability of the rotating hairy black hole. The Gibbs free energy is given by
\begin{eqnarray}
& & \mathcal{G} =\mathcal{M} - T_h S_{BH} \,,
\label{Angularmomentum}
\end{eqnarray}
and its global minimum yields the state of a system for a fixed $(T_h, J)$. In Figures~\ref{TempvsGibbsvsAlphaJPt3LPt4} and \ref{TempvsGibbsvsLJPt3AlphaPt2}, we have shown the behaviour of the Gibbs free energy for different values of $\alpha$ and $\ell$. We see that the smaller black hole branch has a lower free energy compared to the larger black hole branch at all temperatures. This suggests that the smaller black hole branch is indeed thermodynamically more favourable compared to the larger black hole branch. Again the free energy profiles for different values of $\ell$ overlap with each other, suggesting limited thermodynamic effects of this parameter. Moreover, these results again suggest the thermodynamic stability of these hairy black holes is qualitatively similar to the Kerr case.

\begin{figure}[h!]
\begin{minipage}[b]{0.5\linewidth}
\centering
\includegraphics[width=2.8in,height=2.0in]{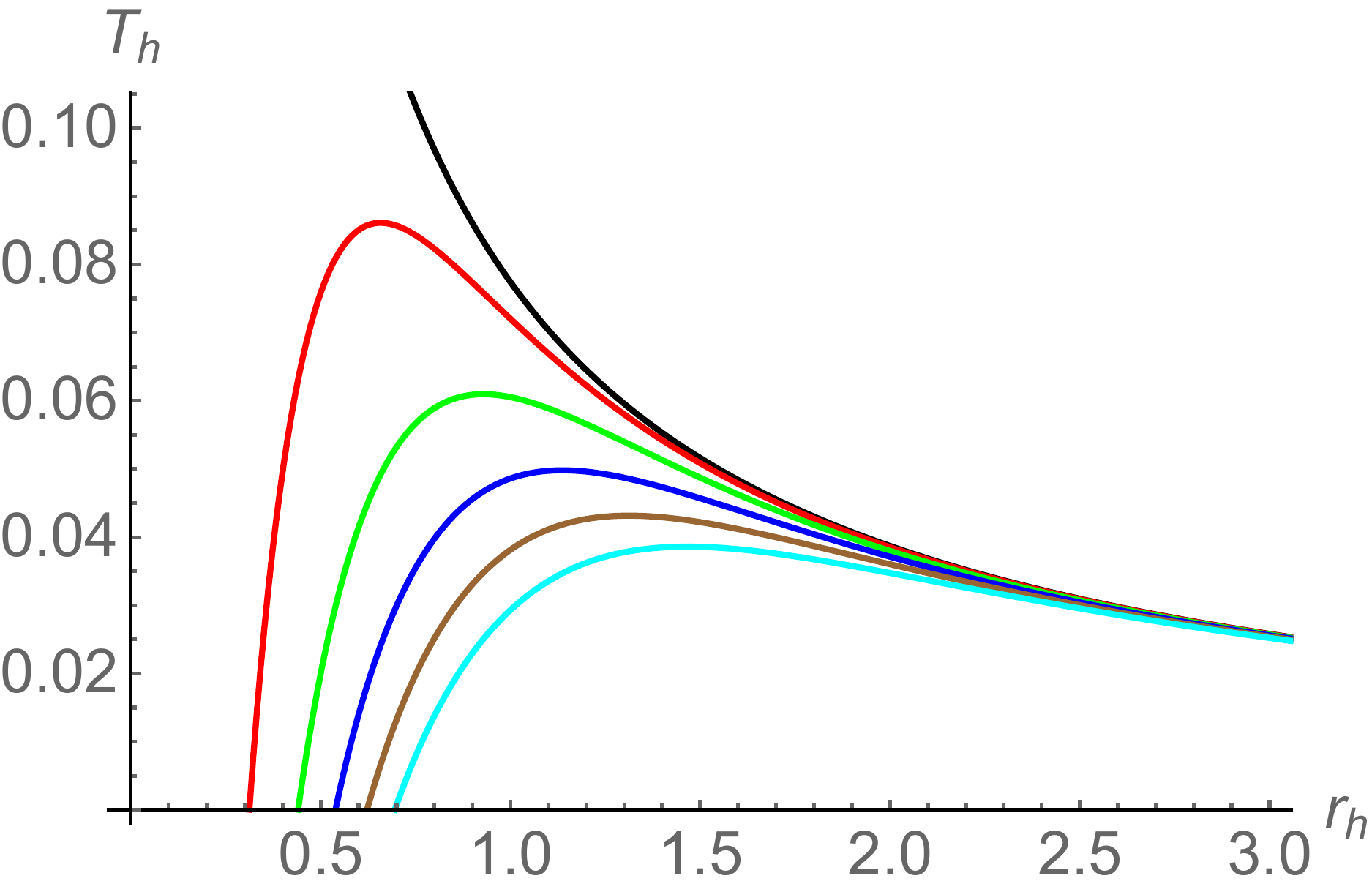}
\caption{ \small The variation of Hawking temperature with respect to the horizon radius for different values of $J$. Here $\alpha=0.2$ and $\ell=0.4$ are used. The black, red, green, blue, brown, and cyan curves correspond to $J=0$, $0.1$, $0.2$, $0.3$, $0.4$, and $0.5$ respectively.}
\label{rhvsTempvsJAlphaPt2LPt4}
\end{minipage}
\hspace{0.4cm}
\begin{minipage}[b]{0.5\linewidth}
\centering
\includegraphics[width=2.8in,height=2.3in]{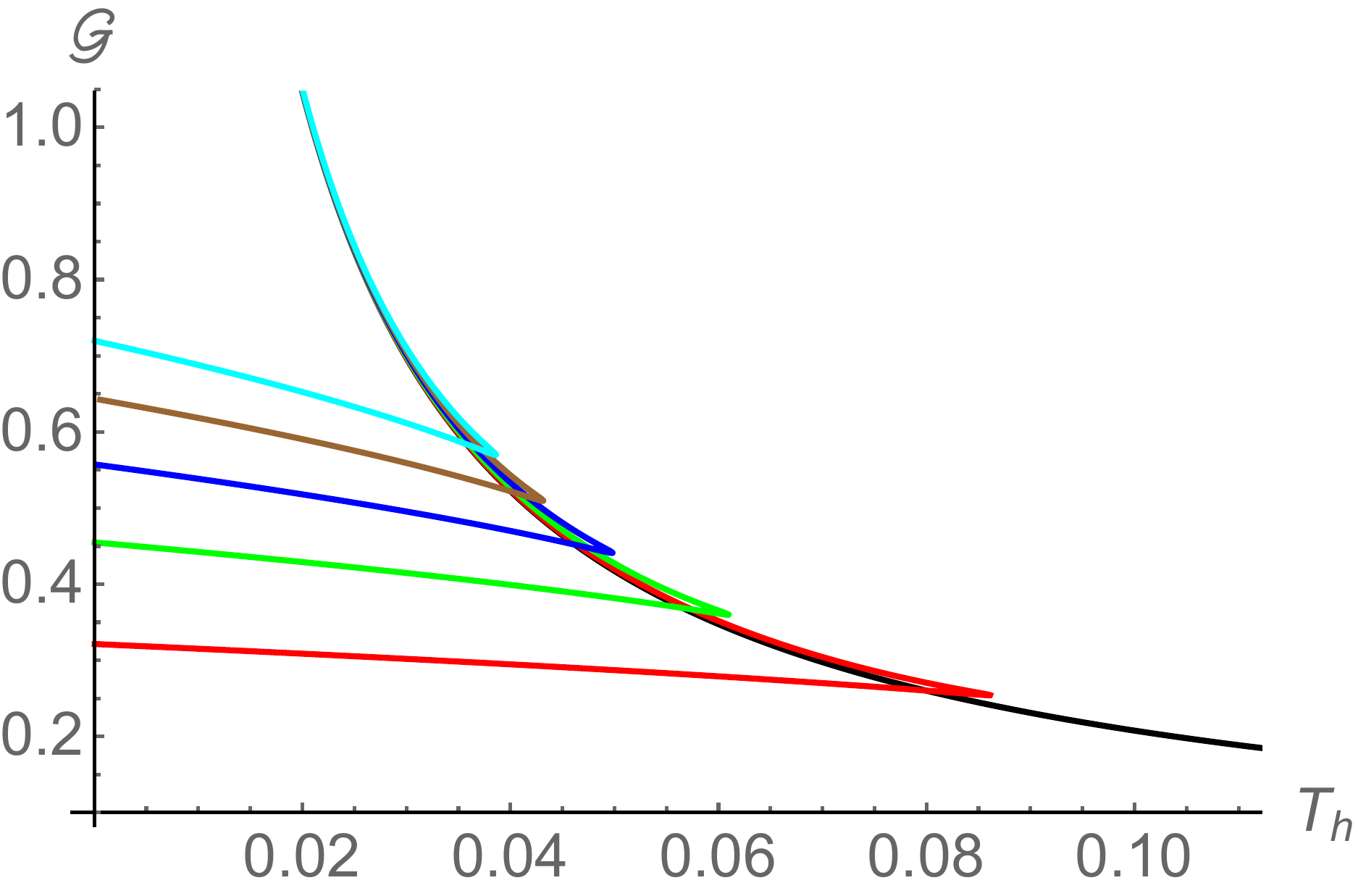}
\caption{\small The variation of free energy $\mathcal{G}$ with respect to the Hawking temperature for different values of $J$.  Here $\alpha=0.2$ and $\ell=0.4$ are used. The black, red, green, blue, brown, and cyan curves correspond to $J=0$, $0.1$, $0.2$, $0.3$, $0.4$, and $0.5$ respectively. In all these curves, the horizon radius $r_h$ increases from left to right and then up.}
\label{TempvsGibbsvsJAlphaPt2LPt4}
\end{minipage}
\end{figure}

To make the thermodynamic analysis more complete, we have also investigated the thermodynamic behaviour of the hairy black holes for different values of angular momentum. These results are shown in Figures~\ref{rhvsTempvsJAlphaPt2LPt4} and \ref{TempvsGibbsvsJAlphaPt2LPt4}. Notice that for the non-rotating case there exists only one black hole branch, the temperature of which decreases with the horizon radius. This behaviour is completely analogous to the Schwarzschild black hole, indicating that static hairy black holes are thermodynamically unstable. With finite angular momentum, a new black hole branch appears for which the temperature increases with the horizon radii. This branch appears for small $r_h$, making the rotating hairy black holes thermodynamically stable up to certain temperatures and radii.

\begin{figure}[h!]
\begin{minipage}[b]{0.5\linewidth}
\centering
\includegraphics[width=2.8in,height=2.0in]{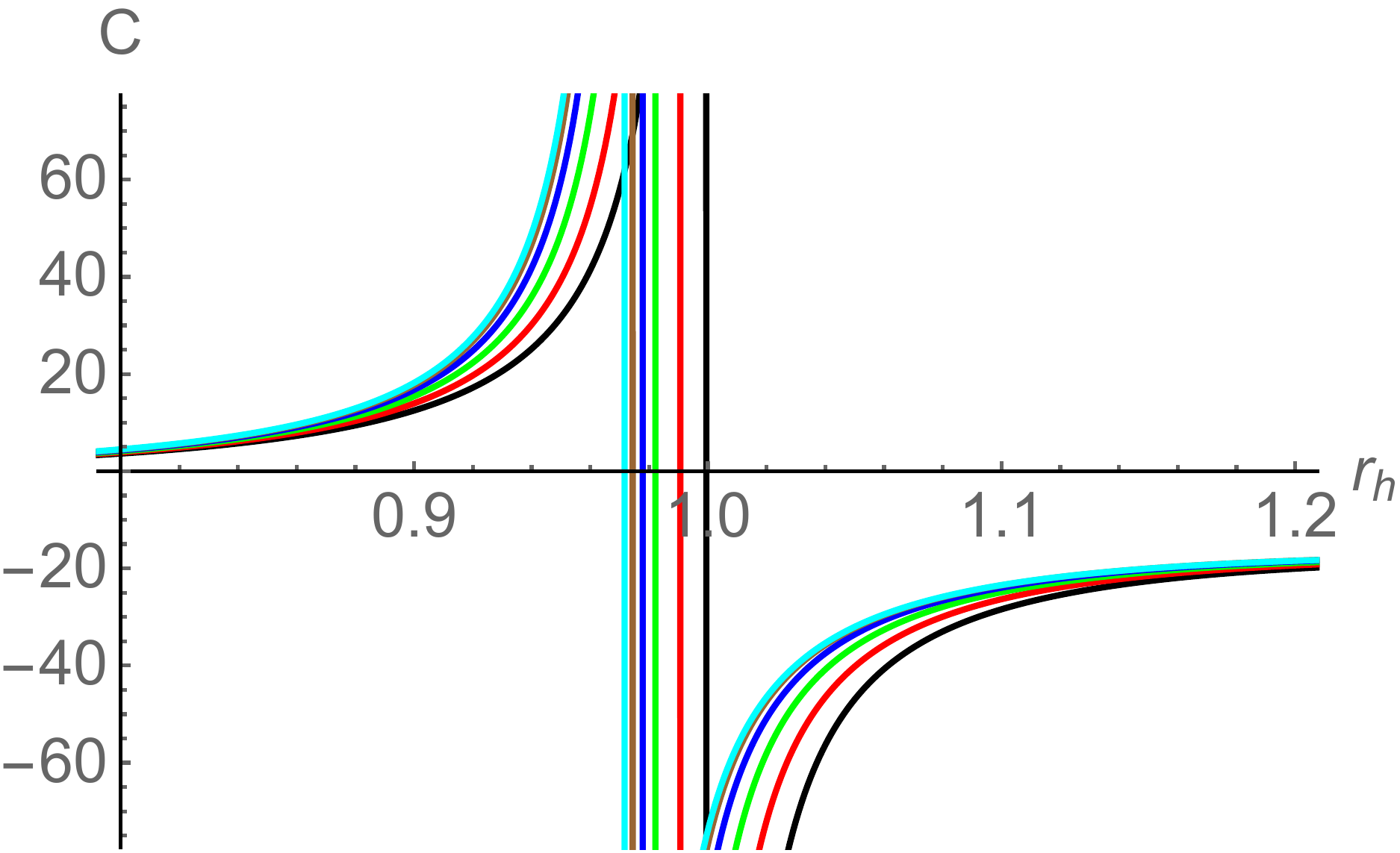}
\caption{ \small The variation of specific heat $C$ with respect to the horizon radius for different values of $\alpha$. Here $J=0.3$ and $\ell=0.4$ are used. The black, red, green, blue, brown, and cyan curves correspond to $\alpha=0$, $0.1$, $0.2$, $0.3$, $0.4$, and $0.5$ respectively.}
\label{rhvsSHvsAlphaJPt3LPt4}
\end{minipage}
\hspace{0.4cm}
\begin{minipage}[b]{0.5\linewidth}
\centering
\includegraphics[width=2.8in,height=2.3in]{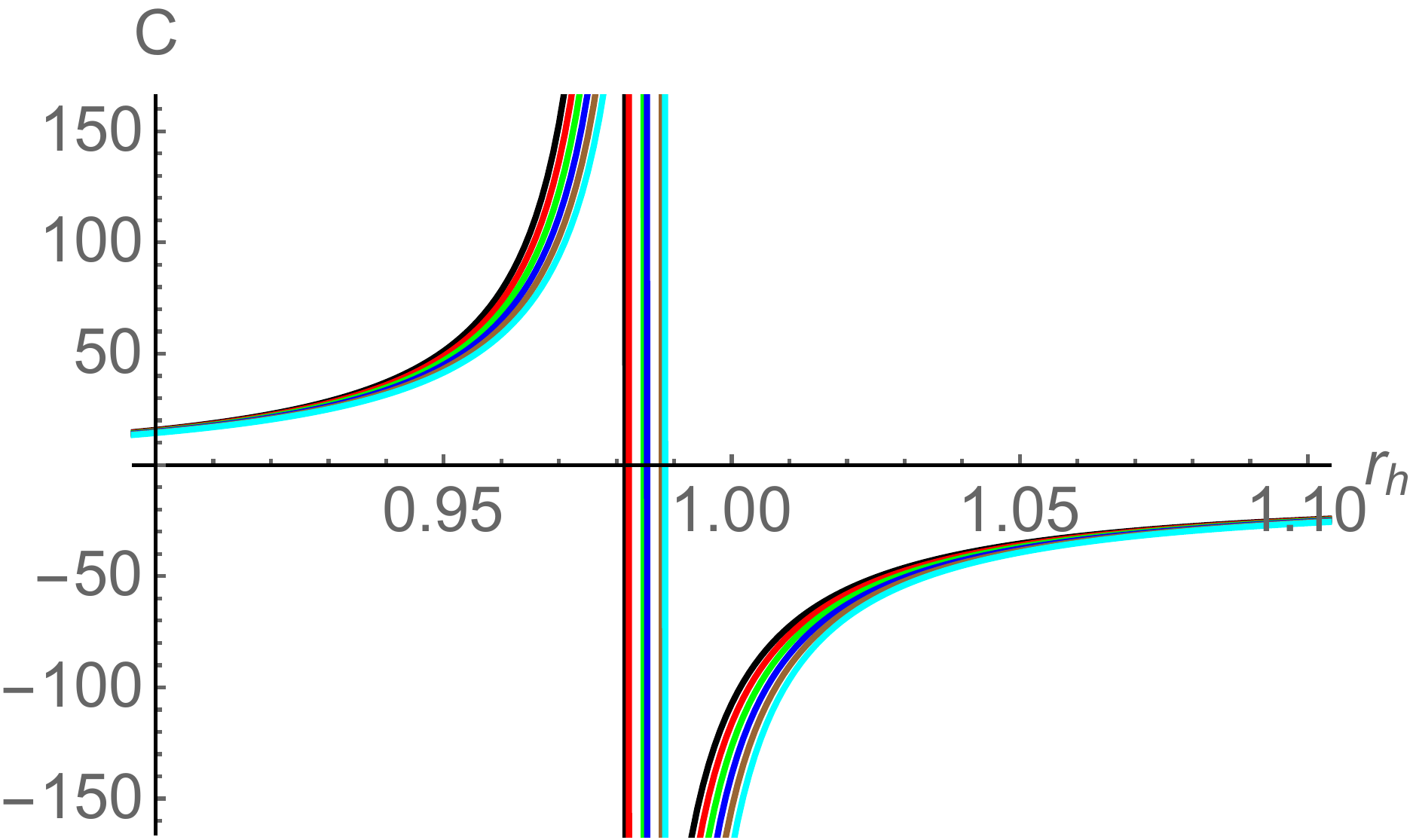}
\caption{\small The variation of specific heat $C$ with respect to the horizon radius for different values of $\ell$. Here $J=0.3$ and $\alpha=0.2$ are used. The black, red, green, blue, brown, and cyan curves correspond to $\ell=0.3$, $0.4$, $0.5$, $0.6$, $0.7$, and $0.8$ respectively.}
\label{rhvsSHvsLJPt3AlphaPt2}
\end{minipage}
\end{figure}

We now discuss the local stability of the rotating hairy black holes. This corresponds to the response of the equilibrium system under a small fluctuation in thermodynamic variables and is quantified by the positivity of the specific heat $C=T_h(\partial S_{BH}/\partial T_h)$. In Figures~\ref{rhvsSHvsAlphaJPt3LPt4} and \ref{rhvsSHvsLJPt3AlphaPt2}, we have shown the behaviour of the specific heat for different values of $\alpha$ and $\ell$. We immediately notice that the specific heat is positive for the small black hole branch, suggesting its thermodynamically stable nature, whereas it is negative for the large black hole branch, suggesting its thermodynamically unstable nature.

\begin{figure}[h!]
\begin{minipage}[b]{0.5\linewidth}
\centering
\includegraphics[width=2.8in,height=2.0in]{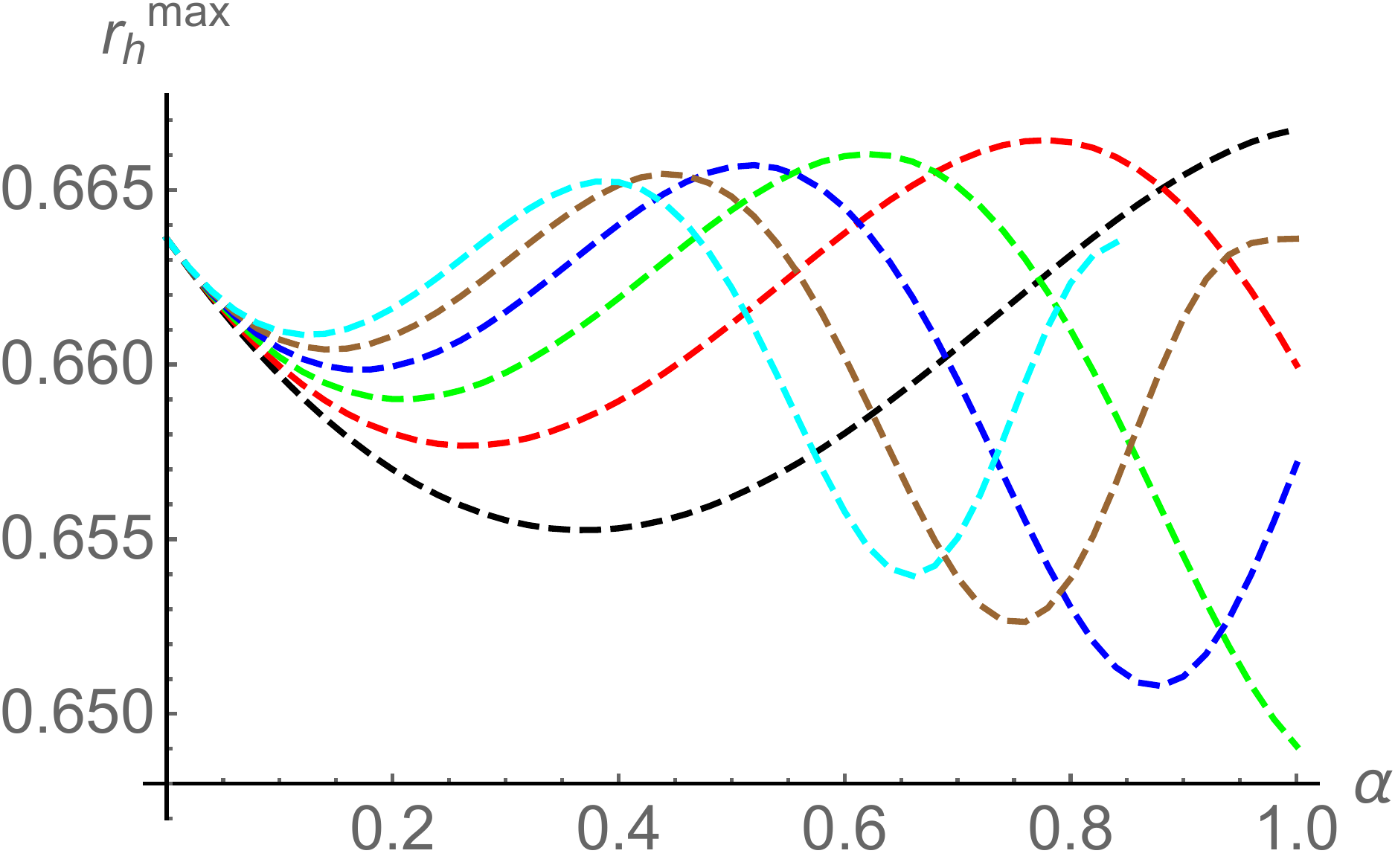}
\caption{ \small The variation of $r_h^{max}$ with respect to $\alpha$ for different values of $\ell$. Here $J=0.1$ is used. The black, red, green, blue, brown, and cyan curves correspond to $\ell=0.3$, $0.4$, $0.5$, $0.6$, $0.7$, and $0.8$ respectively.}
\label{AlphavsrhmaxvsLJPt1}
\end{minipage}
\hspace{0.4cm}
\begin{minipage}[b]{0.5\linewidth}
\centering
\includegraphics[width=2.8in,height=2.3in]{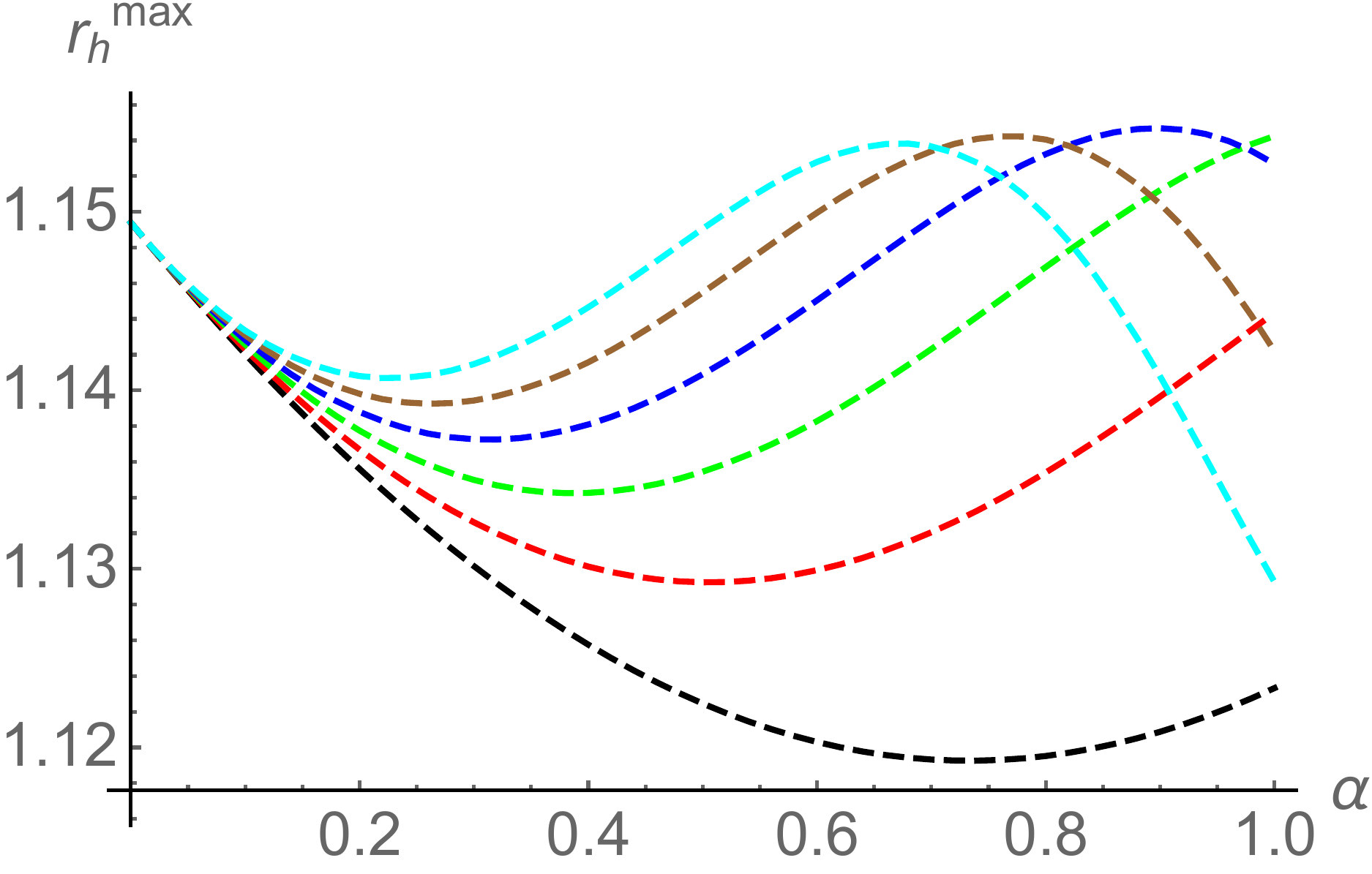}
\caption{\small The variation of $r_h^{max}$ with respect to $\alpha$ for different values of $\ell$. Here $J=0.3$ is used. The black, red, green, blue, brown, and cyan curves correspond to $\ell=0.3$, $0.4$, $0.5$, $0.6$, $0.7$, and $0.8$ respectively.}
\label{AlphavsrhmaxvsLJPt3}
\end{minipage}
\end{figure}
\begin{figure}[h!]
\begin{minipage}[b]{0.5\linewidth}
\centering
\includegraphics[width=2.8in,height=2.0in]{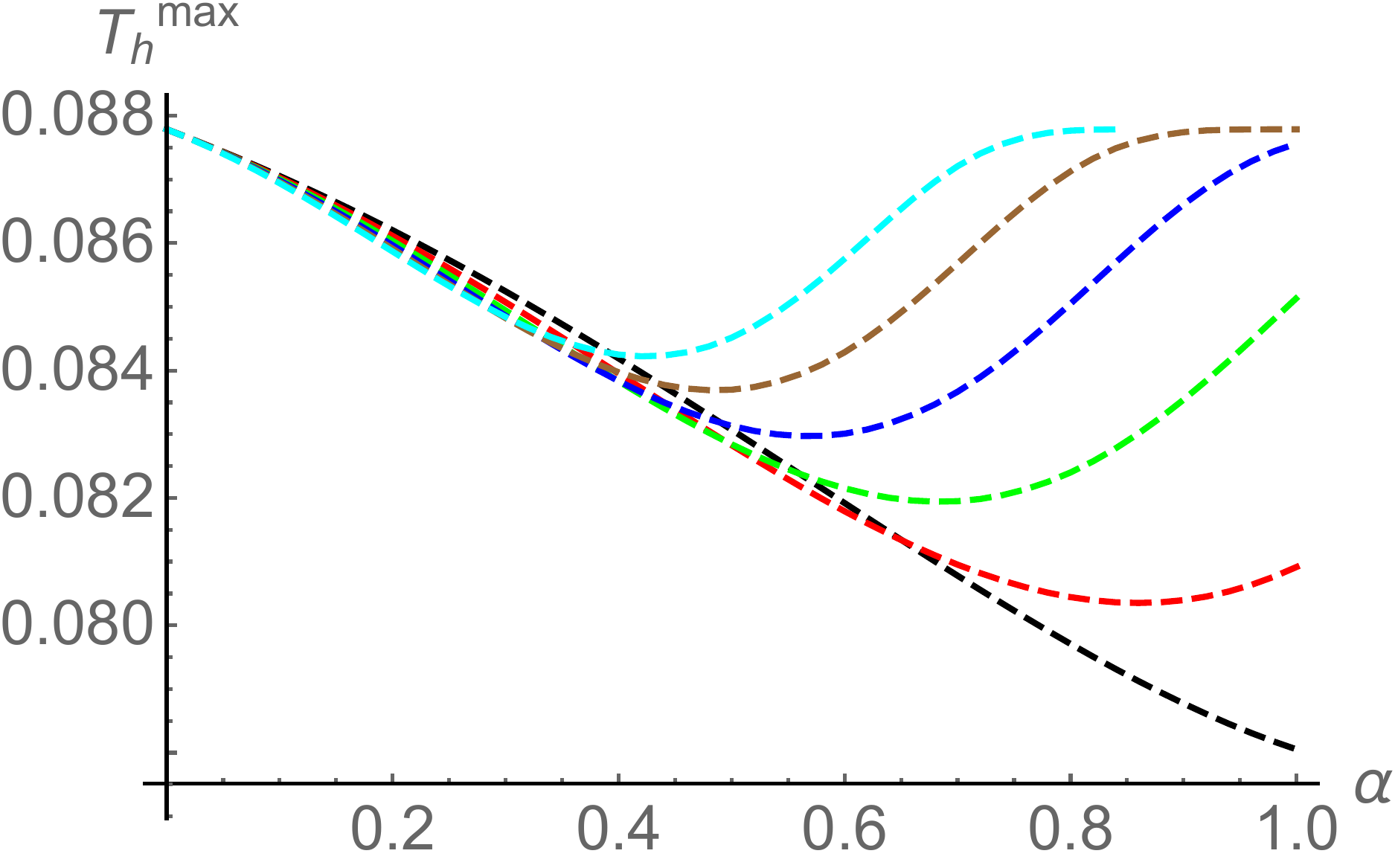}
\caption{ \small The variation of $T_h^{max}$ with respect to $\alpha$ for different values of $\ell$. Here $J=0.1$ is used. The black, red, green, blue, brown, and cyan curves correspond to $\ell=0.3$, $0.4$, $0.5$, $0.6$, $0.7$, and $0.8$ respectively.}
\label{AlphavstempmaxvsLJPt1}
\end{minipage}
\hspace{0.4cm}
\begin{minipage}[b]{0.5\linewidth}
\centering
\includegraphics[width=2.8in,height=2.3in]{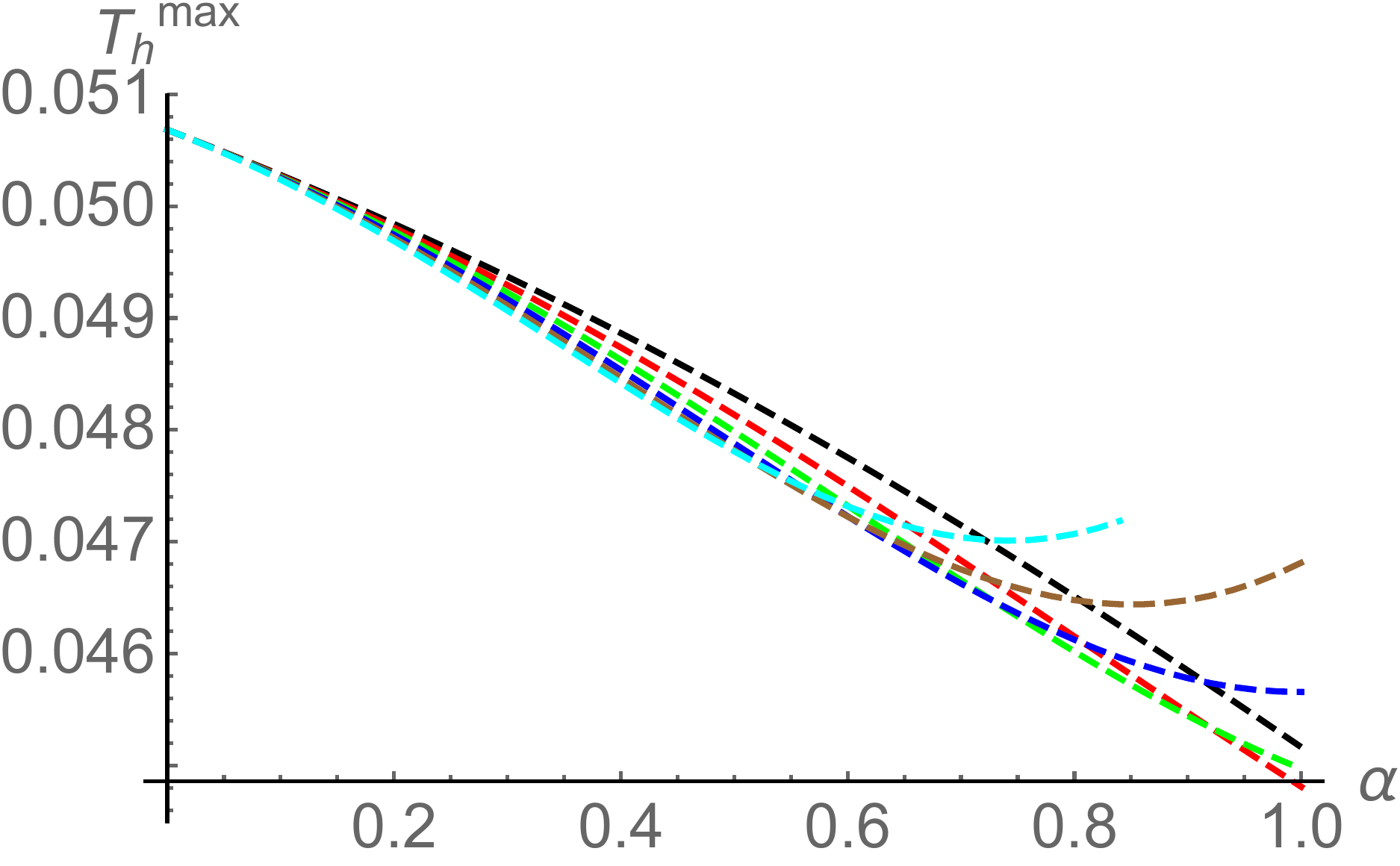}
\caption{\small The variation of $T_h^{max}$ with respect to $\alpha$ for different values of $\ell$. Here $J=0.3$ is used. The black, red, green, blue, brown, and cyan curves correspond to $\ell=0.3$, $0.4$, $0.5$, $0.6$, $0.7$, and $0.8$ respectively.}
\label{AlphavstempmaxvsLJPt3}
\end{minipage}
\end{figure}

It is also interesting to analyse how the horizon radius $r_h=r_h^{max}$, at which the temperature attains a maximum value $T_h^{max}$, varies with the hair parameters. This is important as $r_h^{max}$ specifies the maximum horizon radius around which the stability of black hole changes, particularly, the black holes are stable for $r_h<r_h^{max}$ whereas they are unstable for $r_h>r_h^{max}$.  The results are shown in Figures~\ref{AlphavsrhmaxvsLJPt1} and \ref{AlphavsrhmaxvsLJPt3} for two different angular momenta. We observe that $r_h^{max}$ depends non-trivially on $\alpha$ and $\ell$, and in particular, it can increase or decrease depending on the magnitude of these parameters. This suggests that, compared to the Kerr case, the maximum size of the stable hairy black hole can be smaller or larger depending on $\alpha$ and $\ell$.

One can similarly analyse the dependence of the maximum temperature $T_h^{max}$ on the hairy parameters. The results are shown in Figure~\ref{AlphavstempmaxvsLJPt1} and \ref{AlphavstempmaxvsLJPt3}. We find that $T_h^{max}$ again depends non-trivially on $\alpha$ and $\ell$. In particular, it decreases with $\alpha$ and $\ell$ for relatively small values of $\alpha$ whereas, for higher values of $\alpha$, it can also exhibit an opposite behaviour. Our overall analysis, therefore, suggests that the size, stability, and thermal structure of the hairy black hole depend significantly on the hairy parameters and can be tuned appropriately by suitable choices of these parameters.

We end this section by making some remarks and observations about the first law of thermodynamics like relation for the explored hairy black holes here which we think have not been carefully discussed in the literature \footnote{A related discussion of the first law appeared in \cite{Estrada:2020ptc} for the static hairy black hole.}. There can actually be two different ways by which the first law of black hole thermodynamics can be expressed in the hairy case. As mentioned earlier the shift mass $\mathcal{M}$ is the mass that is observed by an asymptotic observer at infinity. Therefore, one is expected to have $\delta\mathcal{M}$ in the first law of black hole thermodynamics relation, i.e., for the static case
\begin{eqnarray}
& & \delta \mathcal{M}= T_h \delta S_{BH}\,.
\label{firstlaw}
\end{eqnarray}
We find that for the above relation to be true we must treat the shift mass $\mathcal{M}$ as a $r_h$  dependent function and simultaneously must treat $M$ as a $r_h$ independent function. Indeed, we have
\begin{eqnarray}
& & T_h \delta S_{BH} = \frac{1}{2} \left(1+\alpha e^{-\frac{r_h}{M}} \left(1-\frac{r_h}{M} \right)  \right)\delta r_h \,,
\label{firstlaw}
\end{eqnarray}
and this can only be equal to $ \delta \mathcal{M}$, which can be explicitly calculated from Eq.~(\ref{Eq3}), when the parameter $M$ is treated constant of $r_h$. This suggests some kind of Legendre transformation between the thermodynamic functions $M$ and $\mathcal{M}$ when the primary hair $\{\alpha,\ell\}$ is introduced, making the former independent of $r_h$. The same is true for the rotating case as well. Again treating $M$ as a $r_h$ independent function, we can easily verify that the first law of thermodynamic relation
\begin{eqnarray}
& & \delta \mathcal{M}= T_h \delta S_{BH} + \Omega_h \delta J\,.
\label{firstlawrot}
\end{eqnarray}
holds for the rotating hairy black hole case as well.

There can be another interesting scenario by which one can again construct the first law of thermodynamic like relation for the hairy black hole, however now, by treating the thermodynamic function $M$ as a $r_h$ dependent quantity. This can be done by adding another term, corresponding to the primary hair, in the usual first law relation
\begin{eqnarray}
& & \delta \mathcal{M}= T_h \delta S_{BH} + \Omega_h \delta J + \mathcal{Q}_\alpha \delta \mathcal{P}_\alpha  \,,
\label{firstlawmod}
\end{eqnarray}
where $\mathcal{P}_\alpha$ and $ \mathcal{Q}_\alpha$ are now the charge and the corresponding potential associated with the primary hair. Indeed, there have been many suggestions in recent years which strongly advocated for the modification of the standard first law in the presence of additional matter fields. For example, in AdS spaces with an additional scalar field, not only the standard first law gets a correction term [like in Eq.~(\ref{firstlawmod})] due to the scalar field but also the charge and the corresponding potential associated with the primary hair have a well defined meaning \cite{Liu:2013gja,Lu:2014maa,Li:2020spf,Priyadarshinee:2021rch}. For the asymptotically flat black hole considered here, by treating the thermodynamic function $M$ as a $r_h$ dependent function, one particular form of $\mathcal{Q}_\alpha$ and $\mathcal{P}_\alpha$ that would satisfy Eq.~(\ref{firstlawmod}) is
\begin{eqnarray}
& & \mathcal{Q}_\alpha = \frac{\sqrt{\alpha}r_{h}^{4} e^{-\frac{r_h}{\mathcal{M}-\alpha \ell/2}}}{(r_{h}^{2}+a^2)(\mathcal{M}-\alpha \ell/2)^2},~~~~ \mathcal{P}_\alpha = \sqrt{\alpha}  \mathcal{M} \,.
\label{firstlawmod1}
\end{eqnarray}
From the above discussion, it therefore seems that there can be two different formalisms [(\ref{firstlawrot}) or (\ref{firstlawmod})] to think about the first law for the hairy black holes. This raises many important questions such as which one of these two cases is more suitable and accurate in the hairy black hole context, or what are the main differences in the thermodynamic equations of states of the hairy black hole from these formalisms. This is intriguing considering that, unlike the AdS spaces, the additional terms and their physical meaning in the first law due to hair are not explicitly known for the asymptotically flat spaces. These interesting questions are certainly worthy of a separate discussion and we therefore leave this for future work.

\section{Conclusions}
\label{S5}
In this paper, we investigated the axisymmetric analogue of the static hairy black holes which were recently constructed using the gravitational decoupling method. These hairy black holes have additional hairy parameters $\{\alpha, \ell\}$ in the metric and satisfy the strong energy condition (or the dominant energy condition) outside the horizon. We first constructed the rotating counterpart of the hairy solutions using the Newman-Janis and Azreg-A\"{i}nou algorithm and then studied its various properties. We find that all the metric coefficients, except $g_{\theta\theta}$, are modified in the presence of hair, leading to modifications in the horizon structure of the rotating black hole. Similar to the non-hairy Kerr case, there are again two horizons in the rotating hairy case, the smaller one increases whereas the larger one decreases as the strength of the hair $\alpha$ increases. This also leads to the modification in the extremal structure of the hairy black hole, in particular, the mass of the black hole needs to always be greater than the rotation parameter $a$ to avoid the naked singularity. Similarly, we found that the scalar invariants remain finite everywhere outside the horizon and that there are no additional singularities in the hairy case than those already present in the Kerr case.

We then computed various thermodynamic variables and found corrections due to the hair. We found that the thermodynamic structure of the hairy black hole remains qualitatively similar to the Kerr case. In particular, there are again thermodynamically stable small and unstable large black hole branches. We confirmed these results from the free energy and specific heat analysis. We further found that the horizon radius and temperature range for which the rotating hairy black hole remains thermodynamically stable depend non-trivially on the hairy parameters. In particular, they can be smaller/larger compared to the Kerr case depending on the magnitude of $\alpha$ and $\ell$. We also computed quasilocal energy and provided analytic results of it in the slow rotating limit. Finally, we provided some insights into the first law of thermodynamic like relation for these rotating hairy black holes and suggested two different ways by which the first law like relation can be satisfied for the hairy case.

There are many interesting directions to expand this work. The first one would be to investigate the dynamical stability of these black holes against various perturbations. It would also be interesting to verify the correct form of the first law in the hairy case using the Ward formalism \cite{Wald:1993nt,Iyer:1994ys}. It would also be interesting to analyse the fluctuations of the thermodynamic variables and do a comprehensive investigation of the phase transition involved in rotating hairy black holes \cite{Ruppeiner:2007hr,Ruppeiner:2008kd}. On the application side, the obtained rotating hairy solution can also be useful in various cosmological and astrophysical scenarios. In particular, it would be interesting to analyse the imprints of the hairy structure on the black hole shadow, gravitational waves, photon orbits, quasi-normal modes, Hawking radiation etc. and distinguish them from the Kerr case. It would also be interesting the study the effects of cosmological constant and rotational parameter on warped curvatures near the horizon in the lines of \cite{Ovalle:2021jzf,Ovalle:2022eqb}. These analyses can be further used to put bounds on the hairy parameters. We hope to comment on these topics soon.

\section*{Acknowledgments}
The work of S.M.~is supported by the Department of Science and Technology, Government of India under the Grant Agreement number IFA17-PH207 (INSPIRE Faculty Award). The research of I.~B.~is funded by the Start-Up Research
Grant from Science and Engineering Research Board, the Department of Science and Technology, Government of India (Reg. No. SRG/2021/000418).

\appendix
\renewcommand{\theequation}{\thesection.\arabic{equation}}
\addcontentsline{toc}{section}{Appendix}

\section*{Appendix A:  DEC hairy solution and its rotating analogue}
Another static hairy solution that instead satisfies the dominant energy conditions (DEC) $\{\tilde{\rho} \geq |\tilde{p_r}|, \tilde{\rho} \geq |\tilde{p_\theta}|\}$ was further constructed in \cite{Ovalle:2020kpd}. The corresponding metric solution has the form
\begin{eqnarray}
& & e^{g(r)}=e^{-f(r)}=1-\frac{2\mathcal{M}}{r} + \frac{Q^2}{r^2} - \frac{\alpha (\mathcal{M}-\alpha \ell/2) e^{-\frac{r}{\mathcal{M}-\alpha \ell/2}}}{r}\,,
\label{metfunclDEC}
\end{eqnarray}
where again $\mathcal{M}=M+\alpha \ell/2$. The parameters $\ell$ and $Q$ denote the charge that can be associated with the primary hair. It should be noted that some caution is needed in the interpretation of parameter $Q$. In particular, $Q$ does not have to be an electric charge in the usual sense. For example, $Q$ could be the tidal charge of higher dimensional origin or some extra source. However, when $Q$ does represent an electric charge, it means that the electro vacuum or the energy momentum tensor of the Reissner-Nordstr\"{o}m black hole also contains a tensor vacuum. The effective density and pressures are now given by
\begin{eqnarray}
& & \tilde{\rho} = - \tilde{p}_r = \Theta_{0}^{\ 0}=\frac{Q^2}{\kappa^2 r^4} - \frac{\alpha e^{-r/M}}{\kappa^2 r^2} \,,  \nonumber \\
& &  \tilde{p}_\theta = -\Theta_{2}^{\ 2} = \frac{Q^2}{\kappa^2 r^4} - \frac{\alpha e^{-r/M}}{2\kappa^2 M r} \,.
\label{densitypressureDEC}
\end{eqnarray}
from which can see that DEC is readily satisfied in the region $r>2M$ i.e. outside the horizon.

We can again straightforwardly apply the procedure mentioned in Section 3 to find the axisymmetric analogue of (\ref{metfunclDEC}). In this case, the metric coefficients in the Boyer-Lindquist coordinates takes the form
\begin{eqnarray}
& & g_{tt} = \frac{\Delta-a^2\sin^2\theta}{\Sigma}\,,~~~g_{rr}=-\frac{\Sigma}{\Delta}\,,~~~g_{\theta\theta}=-\Sigma \,,\nonumber \\
& & g_{t\varphi} = \frac{a \sin^2\theta (r^2 + a^2 -\Delta)}{\Sigma}\,,~~~ g_{\varphi\varphi} = -\frac{\sin^2\theta \left((r^2 + a^2)^2 -\Delta a^2 \sin^2\theta \right)}{\Sigma} \,, \nonumber \\
& & \Delta(r)= a^2 + r^2 + Q^2 - 2 \mathcal{M} r  - \alpha r \left(\mathcal{M} -\frac{\alpha \ell}{2}  \right) e^{-r/(\mathcal{M} -\frac{\alpha \ell}{2})} \,.
\label{rotatinghairmetDEC}
\end{eqnarray}

\section*{Appendix B:  Newman-Janis algorithm and complexified null tetrads}
The Newman-Janis algorithm involve finding a new null tetrads $V'^{\mu}_{a}$ under the coordinate transformation (\ref{3-9}), such that transformation reduces to the old tetrad and metric when $x'^\mu=\bar{x'}^\mu$, i.e.,
\begin{eqnarray}
V'^{\mu}_{a}(x'^{\sigma}, \bar{x'}^{\sigma})|_{\bar{x'}^{\nu}=x'^{\nu}} = V^{\mu}_{a}(x^{\sigma}) \,,
\label{AppexVtran}
\end{eqnarray}
The effect of this tilde transformation is to generate a new ``tilde'' metric whose components are real functions of complex variable,
\begin{eqnarray}
g_{\mu\nu} \rightarrow g'_{\mu\nu}: x'^{\mu} \times x'^{\nu} \longmapsto \mathbb{R} \,,
\label{Appexgtran}
\end{eqnarray}
A particular transformation of the null tetrad vectors for the hairy metric (\ref{metfunclSEC}) that satisfies the criteria in Eqs.~(\ref{AppexVtran}) and (\ref{Appexgtran}) and give the rotating metric (\ref{rotatinghairmet}) after the transformation (\ref{3-12}) is
\begin{eqnarray}
& & l^{\mu} \rightarrow  l'^{\mu} = \delta^{\mu}_{r'} \,, \nonumber \\
& & n^{\mu} \rightarrow n'^{\mu} = \delta^{\mu}_{u'} - \frac{1}{2} \left(1- \mathcal{M} \left(\frac{1}{r'} + \frac{1}{\bar{r'}} \right) + \frac{\alpha}{4} (r' + \bar{r'})\left(\frac{1}{r'} + \frac{1}{\bar{r'}} \right) e^{\frac{-(r' + \bar{r'})}{2 \mathcal{M} - \alpha \ell}} \right)  \delta^{\mu}_{r'} \,, \nonumber \\
& & m^{\mu} \rightarrow m'^{\mu}  = \frac{1}{\sqrt{2} r'} \left(\delta^{\mu}_{\theta'} + \frac{i}{\sin{\theta'}} \delta^{\mu}_{\phi'}  \right) \,,
\label{tildetetraddef}
\end{eqnarray}
At this point we like to emphasize that the above transformation of the null tetrad vectors is not unique and many similar transformations which satisfy Eqs.~(\ref{AppexVtran}) and (\ref{Appexgtran}) can be constructed. In particular, a certain level of arbitrariness in the method of complexification always appear in the Newman-Janis algorithm. Indeed, most of the time, there is no unique way of complexifying all the terms in the tetrad in the same way. This issue of arbitrary complexification in the Newman-Janis algorithm is known for a long time and has been discussed thoroughly in the literature. In (\ref{tildetetraddef}), the complexification issue is fixed by demanding that the resultant rotational metric using the Newman-Janis algorithm matches to the metric one gets using a without complexification procedure used in section 3.

\bibliography{references}
\bibliographystyle{./utphys1}


\end{document}